\documentclass[aps, prb, preprint, longbibliography, nofootinbib]{revtex4-2}

\usepackage{amsmath}
\usepackage{amssymb}
\usepackage{amsfonts}
\usepackage{graphicx}        
\usepackage{dcolumn}        
\usepackage{bm}             
\usepackage{hyperref}        
\usepackage{cleveref}        
\usepackage{xcolor}          
\usepackage{booktabs}        
\usepackage{microtype}

\hypersetup{
  colorlinks = true,
  linkcolor  = blue,
  citecolor  = blue,
  urlcolor   = blue
}

\begin{document}

\title{Entanglement signature of fully and partially dimerized phases in frustrated spin chains}

\author{Wuttichai Pankeaw}
\email{pankeaw.wutti@gmail.com}

\author{Teparksorn Pengpan}
\email{teparksorn.pengpan@gmail.com}

\author{Pruet Kalasuwan}
\email{pruet.k@psu.ac.th}

\affiliation{Division of Physical Science, Faculty of Science, Prince of Songkla University, Songkhla, 90110, Thailand.}

\begin{abstract}
The von Neumann entanglement entropy of exact valence-bond ground states is studied in two frustrated one-dimensional spin chains: the spin-$\tfrac{1}{2}$ Majumdar--Ghosh (MG) model and the spin-$\tfrac{3}{2}$ $J_1$--$J_2$--$J_3$ chain in its fully dimerized (FD) and partially dimerized (PD) phases. Using matrix-product-state representations, the entropy is computed as a function of system size for three complementary bipartitions—half-chain, single-site, and pairwise—under both open and periodic boundary conditions. In all cases, the entropy saturates to a finite constant in the thermodynamic limit, confirming area-law behavior. The saturation values, extracted via finite-size scaling, are directly related to the underlying virtual-spin bond structure. The MG model and FD phase exhibit similar entanglement behavior, differing primarily in saturation magnitude determined by spin value and bond multiplicity, and both display even--odd oscillations and exponential convergence with system size. In contrast, the PD phase shows qualitatively distinct signatures, including multiple half-chain saturation values depending on the bond type at the cut, asymmetric edge contributions in the single-site entropy, and a multi-band structure in the pairwise entropy reflecting the coexistence of single- and double-singlet bonds. These results establish entanglement entropy as a robust signature of frustrated bond architecture, enabling clear distinction among dimerized phases with different spin magnitude, bond multiplicity, and dimerization patterns.
\end{abstract}


\maketitle

\section{Introduction}
\label{sec:introduction}

Quantum entanglement is a defining feature of many-body quantum systems and has become a central tool for characterizing quantum 
phases and phase transitions~\cite{Osterloh2002,Osborne2002,Vidal2003, Latorre2004,Amico2008}. The von Neumann entanglement 
entropy, defined from the reduced density matrix of a subsystem, provides a quantitative measure of nonlocal quantum correlations 
that often reveals ground-state properties inaccessible to conventional order parameters~\cite{Nielsen2000,Laflorencie2016}. 
Beyond the entropy itself, the full eigenvalue spectrum of the reduced density matrix, known as the entanglement 
spectrum~\cite{LiHaldane2008}, encodes richer information about the underlying quantum state, including edge excitations and 
topological features. In gapped two-dimensional systems with topological order, the entanglement entropy contains a universal 
subleading constant, the topological entanglement entropy, which reflects the long-range quantum structure of the ground 
state~\cite{KitaevPreskill2006,LevinWen2006}. In one-dimensional gapped systems, the entanglement entropy obeys an area law and 
saturates to a finite constant in the thermodynamic limit~\cite{Hastings2007,Eisert2010}, whereas at quantum critical points it 
grows logarithmically with subsystem size, governed by the central charge of the underlying conformal field 
theory~\cite{Calabrese2004,Vidal2003}. These contrasting behaviors establish entanglement entropy as a versatile probe of quantum
phases in low-dimensional systems.

One-dimensional quantum spin chains provide a primary setting for these ideas. Single-site entropy, two-site concurrence, and 
block entanglement entropy have been widely used to detect quantum phase transitions, characterize critical points, and reveal 
ground-state structures in a variety of spin models~\cite{Osterloh2002,Osborne2002,Vidal2003,WangWang2002}. Among these systems, 
integer-spin antiferromagnetic chains are distinguished by a finite excitation gap and exponentially decaying correlations, as 
conjectured by Haldane~\cite{Haldane1983a,Haldane1983b}. The Affleck--Kennedy--Lieb--Tasaki (AKLT) model provides an exactly 
solvable realization of this gapped phase through the valence-bond-solid (VBS) construction~\cite{AKLT1987,AKLT1988}, in 
which each physical spin is decomposed into virtual spin-$\tfrac{1}{2}$ degrees of freedom that form singlet bonds between
neighboring sites. This structure admits a compact representation as a matrix product 
state~\cite{Schollwock2011,Perez2007,Verstraete2008}, which encodes entanglement directly through its virtual bonds and provides 
an analytically tractable framework for studying ground-state properties. Exact entanglement calculations in these VBS states 
show that the entropy saturation value is determined by the virtual-bond structure, with additional boundary contributions 
arising from edge states under open boundary conditions~\cite{Fan2004,Katsura2007,Geraedts2010,Hirano2007}.

Frustration arises when competing interactions or lattice geometry prevent the simultaneous minimization of all bond energies, 
leading to degenerate ground states and unconventional ordering phenomena~\cite{Toulouse1977,Villain1977}. In the sense of 
Toulouse, a plaquette is frustrated when the product of the signs of its exchange interactions is negative, implying that at least 
one bond cannot be satisfied in any spin configuration~\cite{Toulouse1977}. Such frustration may originate from competing 
interactions—for example, ferromagnetic nearest-neighbor and antiferromagnetic next-nearest-neighbor couplings—or from lattice 
geometry, as in triangular or close-packed lattices with antiferromagnetic interactions~\cite{Diep2020,Wannier1950}. Its 
consequences include macroscopic ground-state degeneracy, non-collinear spin structures, and the breakdown of conventional order 
parameters~\cite{Diep2020}. In one-dimensional quantum spin chains, frustration induced by competing interactions can stabilize 
dimerized ground states with nontrivial valence-bond structures. A paradigmatic example is the spin-$\tfrac{1}{2}$ $J_1$--$J_2$ 
chain, whose Majumdar--Ghosh point realizes an exactly solvable fully dimerized state of nearest-neighbor singlets, while 
higher-spin extensions such as the spin-$\tfrac{3}{2}$ $J_1$--$J_2$--$J_3$ model can support both fully and partially 
dimerized phases with distinct bond multiplicities.

These frustrated spin chains provide a natural setting for investigating whether entanglement entropy can probe dimerized phases 
beyond the identification of phase transitions. Previous studies have shown that entanglement entropy and related quantities can 
detect the onset of dimerization and locate phase boundaries in the spin-$\tfrac{1}{2}$ $J_1$--$J_2$ 
chain~\cite{Majumdar1969a,Majumdar1969b,Majumdar1970,Chhajlany2007,Dixit2012,Alet2010}. A complementary and less explored 
question is whether entanglement entropy can resolve the internal bond structure of these phases, namely how valence-bond singlets 
are spatially distributed and combined within a gapped phase away from criticality. This issue becomes particularly relevant 
in higher-spin systems, where multiple singlets may reside on a single bond and qualitatively different dimerization patterns 
can coexist.

For non-frustrated VBS ground states, this connection is well established. Exact results for the spin-1 AKLT chain confirm that 
the entanglement entropy saturates to a value determined entirely by the underlying virtual-spin 
structure~\cite{Fan2004,Katsura2007,Geraedts2010}, demonstrating that bipartite entanglement directly reflects the valence-bond 
architecture. For frustrated higher-spin chains, however, analytical results remain limited. Numerical studies have examined 
entanglement behavior across phase diagrams for spin-$\tfrac{1}{2}$, 1, and $\tfrac{3}{2}$ systems~\cite{Goli2013,Boette2016}, 
and recent work has investigated entanglement scaling in higher-spin VBS states on ladder geometries~\cite{Pankeaw2022}. 
A systematic study comparing entanglement entropy across frustrated dimerized phases and different spin magnitudes is still 
lacking.

The spin-$\tfrac{3}{2}$ $J_1$--$J_2$--$J_3$ chain~\cite{CAM} is particularly well suited to address this problem. It supports both 
a fully dimerized phase, structurally analogous to the Majumdar--Ghosh state but with triple-singlet bonds, and a partially 
dimerized phase characterized by alternating single- and double-singlet bonds. The coexistence of these distinct bond structures 
within a single model makes it an ideal platform for investigating how entanglement encodes frustrated bondarchitecture.

In this work, we compute the von Neumann entanglement entropy of exact valence-bond ground states in the spin-$\tfrac{1}{2}$ 
Majumdar--Ghosh model and the spin-$\tfrac{3}{2}$ $J_1$--$J_2$--$J_3$ chain in both its fully and partially dimerized phases. 
Three complementary bipartitions are considered: the half-chain partition, which probes area-law scaling; the single-site 
partition, which resolves local and edge contributions; and pairwise partitions, which capture distance-dependent correlations. 
Using exact matrix product state representations under both open and periodic boundary conditions, the entanglement entropy is 
evaluated as a function of system size and its saturation values are extracted via finite-size scaling, allowing a direct 
connection to the underlying virtual-bond structure. This approach enables identification of entanglement signatures that are 
universal across frustrated dimerized phases, as well as features that distinguish different spin sectors and uniquely 
characterize partial dimerization.

The remainder of this paper is organized as follows. Section~2 introduces the models and their exact ground states. Section~3 
presents the bipartitions and the entanglement entropy formalism. Section~4 discusses the results, and Section~5 concludes.

\section{Spin systems}
\label{sec:spinsystem}

Present necessary background and related work. Use
\cref{sec:introduction} to cross-reference earlier sections.

\section{Methods}
\label{sec:methods}

In this work, we investigate the behavior of entanglement entropy in two one-dimensional quantum spin systems: the spin-$\tfrac{1}{2}$ Majumdar--Ghosh (MG) $J_1$--$J_2$ chain and the spin-$\tfrac{3}{2}$ $J_1$--$J_2$--$J_3$ chain. The MG model, characterized by antiferromagnetic nearest- and next-nearest-neighbor interactions ($J_1$ and $J_2$), possesses an exactly solvable point at $J_2/J_1 = 1/2$, where the ground state is doubly degenerate and consists of a fully dimerized product of nearest-neighbor singlets \cite{Majumdar1969a}. The spin-$\tfrac{3}{2}$ $J_1$--$J_2$--$J_3$ model extends this framework by including a third-nearest-neighbor interaction $J_3$, supporting fully dimerized ground states and, for suitable couplings, a partially dimerized phase with alternating single and double valence-bond singlets along the $J_1$ bonds \cite{CAM}. These dimerized phases, arising from competing interactions, provide a natural setting for exploring entanglement entropy in both low- and higher-spin chains. Throughout this work, we consider both periodic and open boundary conditions: periodic boundary conditions are imposed for even system sizes ($N = 2$--$48$), while open boundary conditions are employed for odd system sizes ($N = 3$--$49$). In the remainder of this section, we summarize the ground-state structures of these models under these boundary conditions, which serve as reference states for the entanglement entropy analysis.

\subsection{Majumdar--Ghosh model}

\begin{figure}[h] 
\centering
\includegraphics[width=15cm]{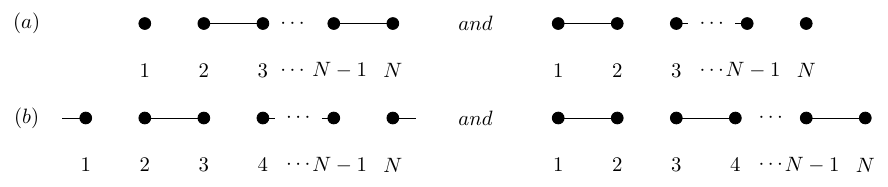}
\caption{The configuration of two-fold degeneracy of the MG ground states, a black dot represents 
spin-$\frac{1}{2}$ at each site and a connected line between sites represents the singlet pair. Where $(a)$ is 
the case of odd $N$ with under OBC, $(b)$ is the case of even $N$ with under PBC.}
\label{fig1}
\end{figure}

The one-dimensional spin-$\tfrac{1}{2}$ $J_1$--$J_2$ antiferromagnetic Heisenberg chain

\begin{equation}
\mathcal{H}_{J_1\text{-}J_2}
  = J_1 \sum_i \mathbf{S}_i \cdot \mathbf{S}_{i+1}
  + J_2 \sum_i \mathbf{S}_i \cdot \mathbf{S}_{i+2},
  \qquad J_{1,2} > 0,
\end{equation}
describes a frustrated system with nearest-neighbour ($J_1$) and next-nearest-neighbour ($J_2$) exchange 
couplings.\cite{Majumdar1969a,Majumdar1969b,Majumdar1970} At the special coupling ratio $J_2/J_1 = 1/2$ 
(equivalently $J_1 = 2J_2$), known as the Majumdar--Ghosh (MG) point, the Hamiltonian can be rewritten, up to an additive 
constant, as a sum of projectors onto total spin-$\tfrac{3}{2}$ on each block of three consecutive sites,\cite{Majumdar1970} which 
guaranties exact, simple ground states. The exact ground states at the MG point are products of nearest-neighbour singlets 
(valence bonds), i.e.\ fully dimerized states \cite{Majumdar1969a,Majumdar1970}. Under open boundary conditions (OBC) with odd $N$ 
[Fig.~\ref{fig1}(a)], the chain ends select a single dimer pattern; however, because an odd-length chain cannot be fully covered 
by nearest-neighbour singlets, the lowest-energy states are more appropriately viewed as a dimer covering on $N-1$ sites together 
with one unpaired spin-$\tfrac{1}{2}$ localized near an edge, naturally interpreted as a domain-wall ``spinon'' between the two 
dimer vacua \cite{Caspers1984}. By contrast, for an even number of spins $N$ under periodic boundary conditions (PBC)
[Fig.~\ref{fig1}(b)], two distinct dimer coverings are possible, with singlets on $(1,2),(3,4),\dots$ or on $(2,3),(4,5),\dots$. 
These coverings are related by a one-site translation and become orthogonal in the thermodynamic limit, yielding a two-fold 
degenerate, gapped ground state that spontaneously breaks lattice translation symmetry \cite{Caspers1984,Kumar2002}.

To explicitly construct the MG ground states, we define a singlet pair formed between site $i$ and its nearest neighbor $i+1$ as

\begin{equation}
\left | \phi \right \rangle_{i,i+1}=\frac{1}{\sqrt{2}}\left ( \left | \uparrow_{i}\downarrow_{i+1} \right 
\rangle - \left | \downarrow_{i}\uparrow_{i+1} \right \rangle \right ).
\end{equation}
Let $\left | s \right \rangle$ represent a single spin-$\frac{1}{2}$ state, where 
$s\in \{\uparrow,\downarrow\}$. For an odd number of spins, the ground state exists under OBC and is given by

\begin{equation}
\left | \psi_{MG}  \right \rangle_{\text{OBC}}=|s\rangle_{1}
  \otimes \bigotimes_{i\in \text{even}}^{N-1} \left | \phi_{\text{even}} 
\right \rangle_{i,i+1}+\bigotimes_{i\in \text{odd}}^{N} \left | \phi_{\text{odd}} \right 
\rangle_{i,i+1} \otimes |s\rangle_{N}.
\end{equation}
For an even number of spins, the ground states under periodic boundary conditions $N+1 \equiv 1$ are given by

\begin{equation}
\left | \psi_{MG}  \right \rangle_{\text{PBC}} = \bigotimes_{i\in \text{even}}^{N}\left | \phi_{\text{even}} 
\right \rangle_{i,i+1}+\bigotimes_{i\in \text{odd}}^{N-1}\left | \phi_{\text{odd}} \right 
\rangle_{i,i+1}.
\end{equation}
Above MG ground states can be depicted in Figure.\ref{fig1}.

\subsection{Dimerized States in Spin-$\frac{3}{2}$ Chain}
For the higher half-integer spin-$S$ chain system, we have considered $S=\frac{3}{2}$ and the 
$J_1$--$J_2$--$J_3$ Hamiltonain governing the system is given by

\begin{equation}
\mathcal{H}_{J_1-J_2-J_3}=J_1\sum_i\bold{S}_i\cdot\bold{S}_{i+1}+J_2\sum_i\bold{S}_{i-1}\cdot 
\bold{S}_{i+1}+J_3\sum_i\left [ \left ( \bold{S}_{i-1}\cdot \bold{S}_{i} \right )\left ( \bold{S}_{i}\cdot 
\bold{S}_{i+1} \right ) +h.c. \right ]
\end{equation}
where $J_1$, $J_2$, and $J_3$ represent the nearest-neighbor, next-nearest-neighbor, and three-site interaction 
strengths, respectively. This Hamiltonian supports fully dimerized ground states while introducing additional 
possibilities, such as partially dimerized phases characterized by alternating strong and weak bonds \cite{CAM}.

\subsubsection{The fully dimerized (FD) ground state}
The fully dimerized (FD) phase in the spin-$\frac{3}{2}$ $J_1$--$J_2$--$J_3$ chain is a gapped, symmetry-broken 
valence bond solid state characterized by three singlets on every other nearest-neighbor $J_1$ bond, forming a 
highly ordered dimer pattern. This phase emerges along the exactly solvable line $J_3 / (J_1 - 2 J_2) = 1/13$ 
for moderate $J_2$, and remains stable for large $J_3$ across all $J_2$ values. It also appears beyond 
critical thresholds such as $J_3/J_1 \approx 0.063$ for $J_2 = 0$, transitioning from critical or partially 
dimerized phases via continuous or first-order transitions \cite{CAM}. The FD phase is considered under both 
periodic and open boundary conditions, which reveal different features in the ground-state degeneracy and edge 
behavior.

We represent each physical spin-$\tfrac{3}{2}$ as the fully symmetric subspace of three virtual 
spin-$\tfrac{1}{2}$ degrees of freedom. For a given physical projection 
$m \in \{-\tfrac{3}{2}, -\tfrac{1}{2}, \tfrac{1}{2}, \tfrac{3}{2}\}$, we introduce an $8\times 1$ column 
vector

\begin{equation}
\label{Tfd}
\begin{split}
T^{[m]}&= \sum_{m=i+j+k} C^{ijk}\,\left | {s^i_1} \right>\otimes \left | {s^j_2}\right>\otimes 
\left |{s^k_3}\right> \\
&=\sum_{m=i+j+k} C^{ijk}|m\rangle,
\end{split}
\end{equation}
where $\left | s \right> \in \bigl\{\left | {\uparrow} \right> = \begin{bmatrix} 1 \\ 0 \end{bmatrix},
\left | {\downarrow} \right> = \begin{bmatrix} 0 \\ 1 \end{bmatrix}\bigr\}$, $i,j,k = \pm\tfrac{1}{2}$, and
$C^{ijk}$ are Clebsch–Gordan coefficients projecting onto the symmetric spin-$\tfrac{3}{2}$ sector.
The four vectors $T^{[m]}$ thus form the columns of an isometry
$T : (\mathbb{C}^2)^{\otimes 3} \to \mathbb{C}^4$ that implements the embedding of the virtual
three–spin space into the physical spin-$\tfrac{3}{2}$ space. 

To form singlet bonds on each virtual leg, we apply the singlet matrix
\begin{equation}
\Phi_{ss} = \frac{1}{\sqrt{2}}
\begin{bmatrix}
0 & 1 \\
-1 & 0
\end{bmatrix}
\end{equation}
to every spin-$\tfrac{1}{2}$ state. This defines modified spin-$\tfrac{3}{2}$ vectors
\begin{equation}
\begin{split}
\tilde{T}^{[m]}&=\sum_{m=i+j+k}C^{ijk} \Phi_{ss}|s^i_1 \rangle \otimes \Phi_{ss}|s^j_2\rangle 
\otimes \Phi_{ss}|s^k_3\rangle \\
       &=\sum_{m=i+j+k} \tilde{C}^{ijk}| m\rangle,
\end{split}
\end{equation}
where the coefficients $\tilde{C}^{ijk}$ are the Clebsch–Gordan coefficients renormalized by the action
of the singlet projectors on the virtual spins. In the MPS language, the local matrices
$\tilde{T}^{[m]}$ provide the matrix representation of the fully dimerized valence–bond state once the
virtual spins on neighboring sites are paired into singlets. The matrix representation of $T^{[m]}$ and $\tilde{T}^{[m]}$ is shown 
in Appendix~\ref{appdxA}.

From these local tensors, we construct the triple–singlet bond state $\left| t \right\rangle_{i,i+1}$,
which forms three singlets between nearest–neighbor sites $i$ and $i+1$, as
\begin{equation}
\left| t \right\rangle_{i,i+1}
= \sum_{m_i, m'_{i+1}} \tilde{T}^{[m_i]\,T}_{i}\,T^{[m'_{i+1}]}_{i+1} \,
   \left| m_i, m'_{i+1} \right\rangle,
\end{equation}
where $\tilde{T}^{[m]\,T}$ denotes the transpose of the vector $\tilde{T}^{[m]}$, and
$\left| m_i, m'_{i+1} \right\rangle$ represents the pair of spin-$\tfrac{3}{2}$ states on sites $i$
and $i+1$. This state realizes a triple–singlet bond on the physical $J_1$ link $(i,i+1)$.

Using the local triple–singlet bond state $\left| t \right\rangle_{i,i+1}$ defined above, we
can now construct the fully dimerized ground states for a chain of length $N$ under different
boundary conditions, in direct analogy with the Majumdar–Ghosh construction (see
Fig.~\ref{fig1}). The structure naturally separates into the cases of even and odd
$N$.

\begin{figure}[h]
  \centering
  \includegraphics[width=0.8\textwidth]{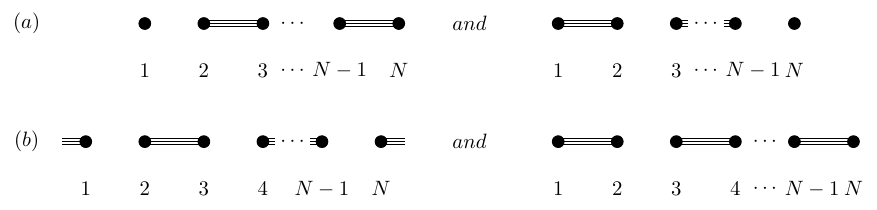}
  \caption{The configurations of the fully dimerized ground states of the spin-$\tfrac{3}{2}$ $J_1$--$J_2$--$J_3$ chain. 
  Black dots denote spin-$\tfrac{3}{2}$ sites and triple lines represent triple-singlet bonds. The two degenerate ground 
  states related by a one-site translation are shown side by side in each case. (a) Odd-$N$ chain under OBC: triple-singlet bonds 
  cover the bulk with one free spin-$\tfrac{3}{2}$ remaining at the left boundary or the right boundary. (b) Even-$N$ chain under 
  PBC.}
  \label{fig2}
\end{figure}

When $N$ is odd, a perfect dimer covering of all sites is impossible. Under open boundary conditions (OBC), 
the fully dimerized pattern can only be realized in the bulk, leaving a free spin-$\tfrac{3}{2}$ degree
of freedom at one edge. There are two such configurations, with the free spin either at the 
left or at the right boundary, as sketched in Fig.~\ref{fig2}(a). The ground state can be written as

\begin{equation}
\left | \psi_{fully}  \right \rangle_{OBC}=\left | m \right \rangle_{1}\otimes 
\bigotimes_{i \in even} ^{N-1} \left | t \right \rangle_{i,i+1} + \bigotimes_{i \in odd} ^{N}\left | t \right 
\rangle_{i,i+1} \otimes \left | m \right \rangle_{N}.
\end{equation}
where $\left| m \right\rangle_{i=1}$ and $\left| m \right\rangle_{i=N}$ denote free edge
spin-$\tfrac{3}{2}$ states at sites $1$ and $N$, respectively.

For even $N$ with periodic boundary conditions (PBC, $N+1 \equiv 1$), there are two exactly
dimerized ground states, corresponding to triple–singlet bonds on either the odd or the even
$J_1$ links as

\begin{equation}
\left | \psi_{fully}  \right \rangle_{PBC}=\bigotimes_{i \in even} ^{N} 
\left| t \right\rangle_{i,i+1} + \bigotimes_{i \in odd} ^{N-1} \left| t \right\rangle_{i,i+1}.
\end{equation}
This configuration is illustrated in Fig.~\ref{fig2}(b); the two covering are related by a one–site
translation and realize the two symmetry–broken fully dimerized ground states of the FD phase.

\subsubsection{The partially dimerized (PD) ground state}

The partially dimerized (PD) phase appears at intermediate frustration and small $J_3$, roughly in the window 
$0.22 \leq J_2/J_1 \leq 0.35$ with $J_3/J_1$ of order $10^{-2}$ ($J_3/J_1 \approx 0.008$ at $J_2/J_1 = 0.3$) 
\cite{CAM}. It is a gapped, translation–symmetry–broken valence–bond phase in which nearest–neighbor $J_1$ bonds
alternate between carrying one and two singlet dimers, making it distinct from the fully dimerized state. In 
the phase diagram it is separated from the surrounding critical phases by a Kosterlitz–Thouless transition and 
from the fully dimerized phase by a first–order line. Under both open and periodic boundary conditions, finite 
chains display characteristic edge physics and degeneracies associated with this pattern of partial 
dimerization.

To formulate the partially dimerized ground states, we use the matrix product state representation. As before, 
each physical spin-$\tfrac{3}{2}$ is built from three virtual spin-$\tfrac{1}{2}$ degrees of freedom, and we 
construct a spin-$\tfrac{3}{2}$ projector $\mathbf{P}_{3/2}$ that projects the virtual space onto the 
symmetric physical subspace. For the PD phase this local projector comes in two inequivalent versions: type $A$ 
(two virtual legs on the left and one on the right) and type $B$ (the opposite configuration). They can be 
written in outer–product form as

\begin{eqnarray}
\mathbf{P}_{3/2}^{A}  & = &  \left | \tfrac{3}{2} \right \rangle\left \langle 1,\tfrac{1}{2} \right | 
+ \left | -\tfrac{3}{2} \right \rangle\left \langle -1,-\tfrac{1}{2} \right |
\nonumber\\
& & + \sqrt{\tfrac{1}{3}} \left | \tfrac{1}{2} \right \rangle\left \langle 1,-\tfrac{1}{2} \right | 
+ \sqrt{\tfrac{2}{3}} \left | \tfrac{1}{2} \right \rangle\left \langle 0,\tfrac{1}{2} \right |
\nonumber \\
& & + \sqrt{\tfrac{1}{3}} \left | -\tfrac{1}{2} \right \rangle\left \langle -1,\tfrac{1}{2} \right | 
+ \sqrt{\tfrac{2}{3}} \left | -\tfrac{1}{2} \right \rangle\left \langle 0,-\tfrac{1}{2} \right | ,
\label{A} \\
\mathbf{P}_{3/2}^{B}  & = &  \left | \tfrac{3}{2} \right \rangle\left \langle \tfrac{1}{2},1 \right | 
+ \left | -\tfrac{3}{2} \right \rangle\left \langle -\tfrac{1}{2},-1 \right |
\nonumber\\
& & + \sqrt{\tfrac{1}{3}} \left | \tfrac{1}{2} \right \rangle\left \langle -\tfrac{1}{2},1 \right | 
+ \sqrt{\tfrac{2}{3}} \left | \tfrac{1}{2} \right \rangle\left \langle \tfrac{1}{2},0 \right |
\nonumber \\
& & + \sqrt{\tfrac{1}{3}} \left | -\tfrac{1}{2} \right \rangle\left \langle \tfrac{1}{2},-1 \right | 
+ \sqrt{\tfrac{2}{3}} \left | -\tfrac{1}{2} \right \rangle\left \langle -\tfrac{1}{2},0 \right | .
\label{B}
\end{eqnarray}
Here the bras $\langle 1,\tfrac{1}{2}|$, $\langle 0,-\tfrac{1}{2}|$, etc.\ denote composite virtual states 
with a “left” and a “right” leg. In type $A$ the virtual configuration is $\langle s^{1}, s^{1/2} | = 
\langle s^{1} |_L \otimes \langle s^{1/2} |_r \equiv \langle L,r |$, while in type $B$ it is 
$\langle s^{1/2}, s^{1} | = \langle s^{1/2} |_l \otimes \langle s^{1} |_R \equiv \langle l,R |$. Collecting 
the Clebsch–Gordan coefficients in matrix form, we obtain local MPS tensors $A^{[m]}$ and $B^{[m]}$, where 
$A^{[m]}$ is a $4\times 2$ matrix with indices $(L,r)$ and $B^{[m]}$ is a $2\times 4$ matrix with indices 
$(l,R)$; the physical index $m = \pm \tfrac{3}{2}, \pm \tfrac{1}{2}$ labels the spin-$\tfrac{3}{2}$ states.

To form singlet bonds along the chain in the partially dimerized pattern, we dress these projectors with the 
spin-$\tfrac{1}{2}$ singlet matrix $\Phi_{ss}$ on the appropriate virtual legs. This defines modified tensors

\begin{align}
\tilde{A}^{[m]} &= A^{[m]} \,\Phi_{ss}, \\
\tilde{B}^{[m]} &= B^{[m]} \,\bigl( \Phi_{ss} \otimes \Phi_{ss} \bigr),
\end{align}
where $\Phi_{ss}$ contracts one virtual spin-$\tfrac{1}{2}$ leg for $\tilde{A}^{[m]}$ and two 
legs for $\tilde{B}^{[m]}$, reflecting the alternating pattern of single and double singlet
bonds. These modified tensors are then arranged in an alternating $A$–$B$ pattern along the
chain to build the MPS representation of the partially dimerized ground state; for odd $N$
under OBC one must additionally distinguish ``double'' and ``single'' boundary configurations,
depending on whether the left edge starts with a doubly or singly dimerized bond.

\begin{figure}[h] 
\centering
\includegraphics[width=15cm]{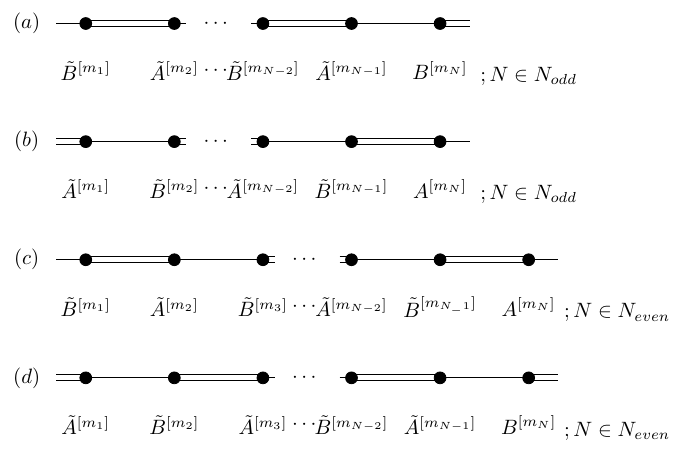}
\caption{Schematic MPS representations of the partially dimerized spin-$\tfrac{3}{2}$ chain. Black dots denote 
physical spin-$\tfrac{3}{2}$ sites, and lines represent singlet bonds between virtual spin-$\tfrac{1}{2}$ 
degrees of freedom. (a),(b) Odd $N$: the left and right virtual legs are unmatched (single vs.\ double), 
corresponding to free edge spins under OBC. (c),(d) Even $N$: the boundary virtual legs are matched and can be 
contracted into a closed loop, corresponding to PBC.}
\label{fig3}
\end{figure}

The corresponding MPS ground states are given by sums over all physical spin indices $m_i = \pm \tfrac{3}{2}, 
\pm \tfrac{1}{2}$. For odd $N$ under OBC, we distinguish the “single” and “double” boundary configurations, 
which read

\begin{subequations}
\begin{align}
\left | \psi_{PD}  \right \rangle_{\text{odd, single}} &= 
\sum_{m_1, \ldots, m_N}\tilde{B}^{[m_1]}\tilde{A}^{[m_2]}\ldots\tilde{B}^{[m_{N-2}]} 
\tilde{A}^{[m_{N-1}]}B^{[m_N]}\left | m_1,m_2,\ldots,m_N \right \rangle, \\
\left | \psi_{PD}  \right \rangle_{\text{odd, double}} &= 
\sum_{m_1, \ldots, m_N}\tilde{A}^{[m_1]}\tilde{B}^{[m_2]}\ldots\tilde{A}^{[m_{N-2}]} 
\tilde{B}^{[m_{N-1}]}A^{[m_N]}\left | m_1,m_2,\ldots,m_N \right \rangle.
\end{align}
\end{subequations}
These sums run over all spin-$\tfrac{3}{2}$ configurations; only those terms for which the matrix products are 
nonzero contribute. The corresponding schematics are shown in Fig.~\ref{fig3}(a) and (b), respectively.

For even $N$, partially dimerized (PD) ground states arise under periodic boundary condition. In this case, 
the first and last spins are paired through a singlet bond, and the virtual indices are closed by taking a 
trace. The PD ground state can be written as

\begin{equation}
\left | \psi_{PD}  \right \rangle^{PBC}_{\text{even}} =\sum_{m_1, \ldots, m_N} 
tr\left ( \tilde{A}^{[m_1]}\tilde{B}^{[m_2]}\ldots \tilde{A}^{[m_{N-2}]}\tilde{B}^{[m_{N-1}]} 
\tilde{B}^{[m_N]} \right )\left | m_1,m_2,\ldots,m_N \right \rangle,
\end{equation}
where \(\operatorname{tr}(\cdot)\) denotes the contraction over virtual indices, thereby enforcing periodic 
closure by identifying the first and last virtual legs. The corresponding schematic MPS representations for 
even \(N\) under PBC are shown in Fig.~\ref{fig3}(c) and (d).

\section{Entanglement entropy}
\label{sec:entanglemententropy}

The von Neumann entanglement entropy is used to quantify the quantum correlations in a bipartite system 
composed of subsystems $A$ and $B$. For a pure ground state $|\psi\rangle$, the reduced density matrix of 
subsystem $A$ is obtained by tracing out the degrees of freedom of subsystem $B$,
\begin{equation}
    \rho_A = \mathrm{Tr}_B\,|\psi\rangle\langle\psi|,
\end{equation}
and the entanglement entropy is defined as
\begin{equation}
    S_A = -\mathrm{Tr}(\rho_A \ln \rho_A)
        = -\sum_i \lambda_i \ln \lambda_i,
\label{eq:vN}
\end{equation}
where $\lambda_i$ are the eigenvalues of $\rho_A$~\cite{Nielsen2000}.

In this work, we compute $S_A$ for three distinct choices of subsystem $A$, each designed to probe a different 
aspect of the entanglement structure of the ground state.

The first is the half-chain bipartition, in which subsystem $A$ consists of $\lfloor N/2 \rfloor$ contiguous 
spins. We study how $S(N/2)$ varies with system size $N$ to characterize the scaling behavior of entanglement. 
For a gapped system obeying the area law~\cite{Hastings2007, Eisert2010}, the entropy saturates to a finite 
constant,

\begin{equation}
    S(N/2) \xrightarrow{N\to\infty} S_\infty < \infty,
\label{eq:arealaw}
\end{equation}
independent of system size. In the presence of dimerization or geometric frustration, $S(N/2)$ 
may additionally exhibit oscillations between odd and even as a function of $N/2$, reflecting the 
period-2 structure of the underlying lattice. In such cases, we analyze the even and odd subseries 
of $S(N/2)$ separately to isolate the bulk scaling behavior from these finite-size effects. Should 
both subseries converge to the same constant in the large $N$ limit, this is consistent with area 
law saturation~\cite{Hastings2007, Eisert2010}. Furthermore, the saturation value $S_\infty$ is 
not arbitrary: for a valence bond solid (VBS) state with Schmidt rank $\mathcal{D}$ across the 
boundary cut, one expects~\cite{Eisert2010}
\begin{equation}
    S_\infty = 2\ln\mathcal{D},
\label{eq:sinf}
\end{equation}
so that distinct gapped phases may in principle be distinguished by their characteristic values of $S_\infty$.

\begin{figure}[h]
    \centering
    \includegraphics[width=10cm]{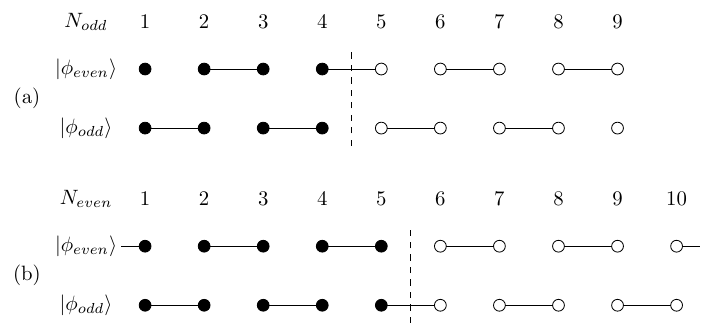}
    \caption{The way to cut half-chain bipartition where (a) $N$ is odd, cutting at $\frac{N-1}{2}$ and (b) 
    $N$ is even, cutting at $\frac{N}{2}$. The dashed lines indicate the bipartition boundaries separating 
    subsystem $A$ (left) from subsystem $B$ (right). The states $\left | \phi_{even} \right \rangle$ and 
    $\left | \phi_{odd} \right \rangle$ denote the two dimer configurations with singlets formed on even and 
    odd bonds, respectively.}
    \label{fig4}
\end{figure}

The second bipartition is the single-site entropy, in which subsystem $A$ consists of a single spin at site $i$, where 
$i \in \left\{ 1,2, \ldots, N\right\}$. The resulting entropy $S(i)$ measures the local entanglement between site 
$i$ and the rest of the chain, and serves as a site-resolved probe of the spatial distribution of quantum correlations across the 
chain~\cite{Osterloh2002}. In particular, $S(i)$ allows us to distinguish bulk sites from boundary sites and to detect any 
inhomogeneity in the entanglement structure arising from edge effects or symmetry breaking.

The third bipartition is the pairwise entropy, in which subsystem $A$ consists of any two sites $i$ and $j$ 
embedded in the chain. The resulting entropy $S(i,j)$ quantifies the entanglement between the pair $\{i,j\}$ 
and the remainder of the chain. Because the distance definition and the available pair types depend on the 
boundary condition, we treat the two cases separately.
 
\begin{figure}[h]
\centering
\includegraphics[width=0.6\textwidth]{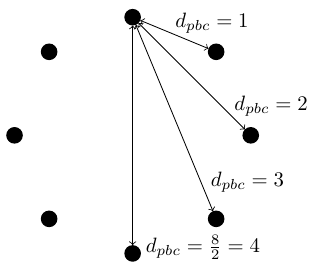}
\caption{The pairwise distance under PBC, $d_{pbc}$ is the translational and reflection symmetry of the 
periodic chain, all pairs at the same chord distance $d_{pbc}$ are equivalent. For example $N=4$, we have 
$d_{pbc} = 1, 2, 3, 4$.}
\label{fig5}
\end{figure}

Under PBC with even $N$, the chain is translationally symmetric and the pairwise entropy depends only on the 
chord distance between the two sites. Due to this symmetry, the distance is defined as
\begin{equation}
    d_{pbc} = \min(|i-j|,\, N - |i-j|),
    \qquad d_{pbc} = 1, 2, \ldots, \tfrac{N}{2},
    \label{eq:dpbc}
\end{equation}
so that $d_{pbc}$ ranges from $1$ (nearest-neighbor pair) to $N/2$
(maximally separated pair). Figure~\ref{fig5} illustrates this
for a chain of $N = 8$ sites, showing the four distinct distances
$d_{pbc} = 1, 2, 3, 4 = N/2$.
 
\begin{figure}[h]
\centering
\includegraphics[width=0.6\textwidth]{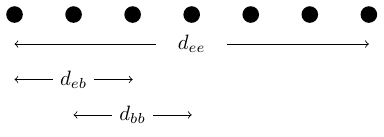}
\caption{The pairwise distance under OBC can be classified into three types. First, the edge-edge distance 
$d_{ee} = N-1$ connects the two boundary spins. Second, the edge-bulk distance $d_{eb} = 1, 2, \ldots, N-2$ 
connects the left edge spin to each bulk spin. Last, the bulk-bulk distance $d_{bb} = 1, 2, \ldots, N-3$ 
connects any two interior bulk spins.}
\label{fig6}
\end{figure}

Under OBC with odd $N$, the chain has a free left edge spin and a free right edge spin, breaking translational 
symmetry. The pairwise entropy $S(i,j)$ depends on both the distance $d = |i-j|$ and the positions of the two 
sites relative to the boundaries. We therefore classify all pairs into three geometrically distinct groups, as 
illustrated in
Fig.~\ref{fig6}:
\begin{itemize}
    \item \textit{Edge--edge pair} ($d_{ee}$): the left edge spin paired with the right edge spin, with fixed 
    distance $d_{ee} = N - 1$. This is a single pair for each $N$.
    \item \textit{Edge--bulk pairs} ($d_{eb}$): the left edge spin paired with any bulk spin, with distance 
    $d_{eb} = 1, 2, \ldots, N-2$.
    \item \textit{Bulk--bulk pairs} ($d_{bb}$): any two bulk spins paired together, with distance $d_{bb} = 1, 
    2, \ldots, N-3$.
\end{itemize}
 
\section{Results and discussion}
\label{sec:resultdiscussion}

We present the von Neumann entanglement entropy computed across three bipartitions --- half-chain, 
single-site, and pairwise --- for the ground states of the spin-$\tfrac{1}{2}$ Majumdar--Ghosh (MG) model and 
the spin-$\tfrac{3}{2}$ $J_1$--$J_2$--$J_3$ chain in both its fully dimerized (FD) and partially dimerized 
(PD) phases. In every case the entropy saturates to a finite, system size independent value in the 
thermodynamic limit, consistent with the area law for gapped one-dimensional 
systems~\cite{Hastings2007,Eisert2010}. The saturation values are determined by the Schmidt rank $\mathcal{D}$ 
across the bipartition cut through $S_\infty = 2\ln\mathcal{D}$~\cite{Eisert2010}, providing a direct 
fingerprint of the underlying valence-bond structure. Each subsection below treats one bipartition type and 
presents the results for all three phases side by side.

\subsection{Half-Chain Entanglement Entropy}

In gapped one-dimensional systems, the entanglement entropy of a finite subsystem approaches its thermodynamic saturation value 
exponentially with subsystem size, with a decay rate governed by the bulk correlation length~\cite{Hastings2007,Eisert2010}. 
To confirm that this behavior applies to the exact valence-bond ground states considered here, and to exclude the logarithmic scaling
associated with critical systems~\cite{Calabrese2004,Vidal2003}, we test both forms using the Majumdar--Ghosh model under periodic
boundary conditions as a representative case. The results, presented in Appendix~\ref{appdxB}, show that the exponential
form provides a significantly better description of the data than the logarithmic form. Accordingly, each sub-series of the half-chain 
entropy is fitted to the exponential form

\begin{equation}
    S = A - B\,e^{-C\,\ell},
    \label{eq:fit_form}
\end{equation}
where $\ell = N/2$ (PBC) or $\ell = (N-1)/2$ (OBC), $A = S_\infty$ is the thermodynamic saturation value, $B$ controls the amplitude 
of the finite-size correction, and $C = 1/\xi$ corresponds to the inverse correlation length governing the convergence. The resulting 
fit parameters are summarized in Tables~\ref{tab:mg_halfchain_fit}--\ref{tab:pd_obc_fit}.

\subsubsection{Majumdar--Ghosh Model}

\begin{figure}[h]
\centering
\includegraphics[width=\textwidth]{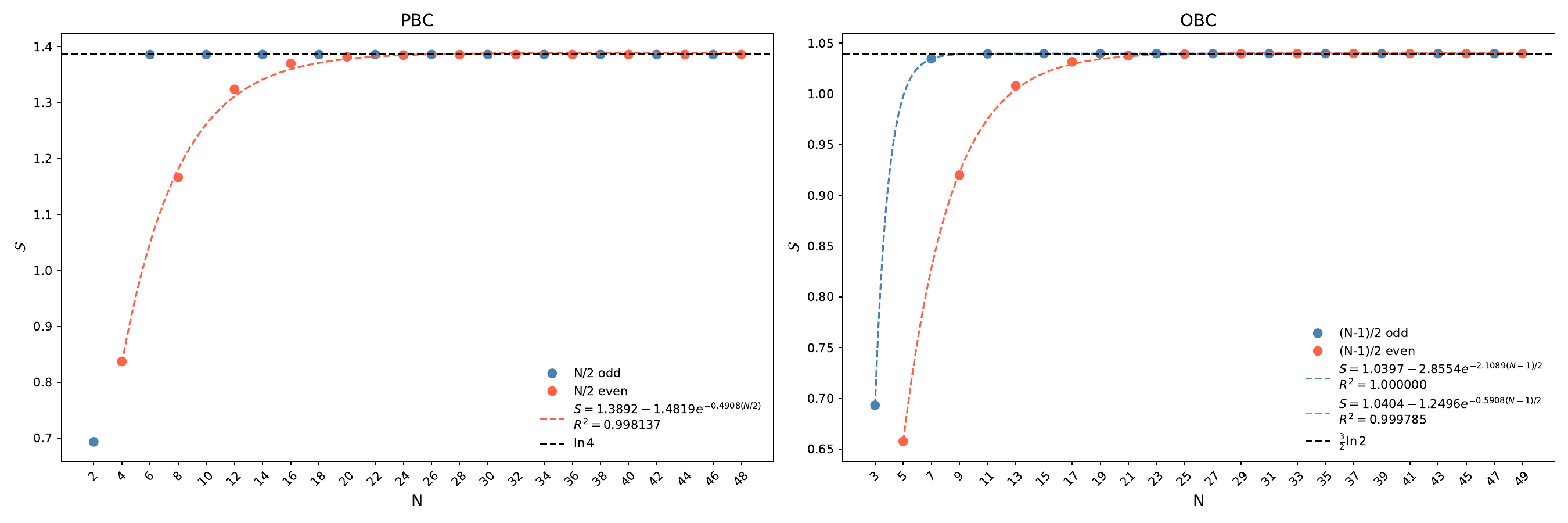}
\caption{Half-chain entanglement entropy of the MG model as a function of system size $N$. Left: even $N$ 
under PBC, with the $N/2$-odd sub-series (blue) sitting exactly on $\ln 4 = 2\ln 2$ and the $N/2$-even 
sub-series (orange) fitted to Eq.~\eqref{eq:fit_form}. Right: odd $N$ under OBC, with both sub-series fitted 
to Eq.~\eqref{eq:fit_form} and converging to $\tfrac{3}{2}\ln 2$ (dashed line).}
\label{fig7}
\end{figure}

Figure~\ref{fig7} shows the half-chain entanglement entropy of the MG model under both boundary 
conditions. Under PBC the $N/2$-odd sub-series attains $S_\infty = 2\ln 2$ exactly for all accessible $N$ (no 
fit needed), while the $N/2$-even sub-series converges from below. Under OBC both sub-series saturate to 
$\tfrac{3}{2}\ln 2$, which is suppressed relative to $2\ln 2$ by the free edge spin-$\tfrac{1}{2}$ under odd 
$N$~\cite{Caspers1984}, contributing only $\tfrac{1}{2}\ln 2$ rather than a full $\ln 2$ to the cut. The 
saturation $2\ln 2$ under PBC corresponds to Schmidt rank $\mathcal{D}=2$, consistent with one nearest 
neighbor singlet crossing the cut~\cite{Majumdar1969a,Majumdar1970}.

\begin{table}[h]
\centering
\caption{Fit parameters of Eq.~\eqref{eq:fit_form} for the MG model.}
\label{tab:mg_halfchain_fit}
\begin{tabular}{llcccc}
\hline\hline
BC & Sub-series & $S_\infty$ & $B$ & $C$ & $R^2$ \\
\hline
PBC & $N/2$ even & $2\ln 2 \approx 1.3892$ & $1.4819$ & $0.4908$ & $0.9981$ \\
PBC & $N/2$ odd & $2\ln 2 \approx 1.3863$ & \multicolumn{3}{c}{exact} \\
\hline
OBC & $(N-1)/2$ odd  & $\tfrac{3}{2}\ln 2 \approx 1.0397$ & $2.8554$ & $2.1089$ & $\approx 1.0000$ \\
OBC & $(N-1)/2$ even & $\tfrac{3}{2}\ln 2 \approx 1.0404$ & $1.2496$ & $0.5908$ & $0.9998$ \\
\hline\hline
\end{tabular}
\end{table}

\subsubsection{Spin-$\tfrac{3}{2}$ $J_1$--$J_2$--$J_3$ Chain}

\paragraph{Fully dimerized phase.}
 
\begin{figure}[h]
\centering
\includegraphics[width=\textwidth]{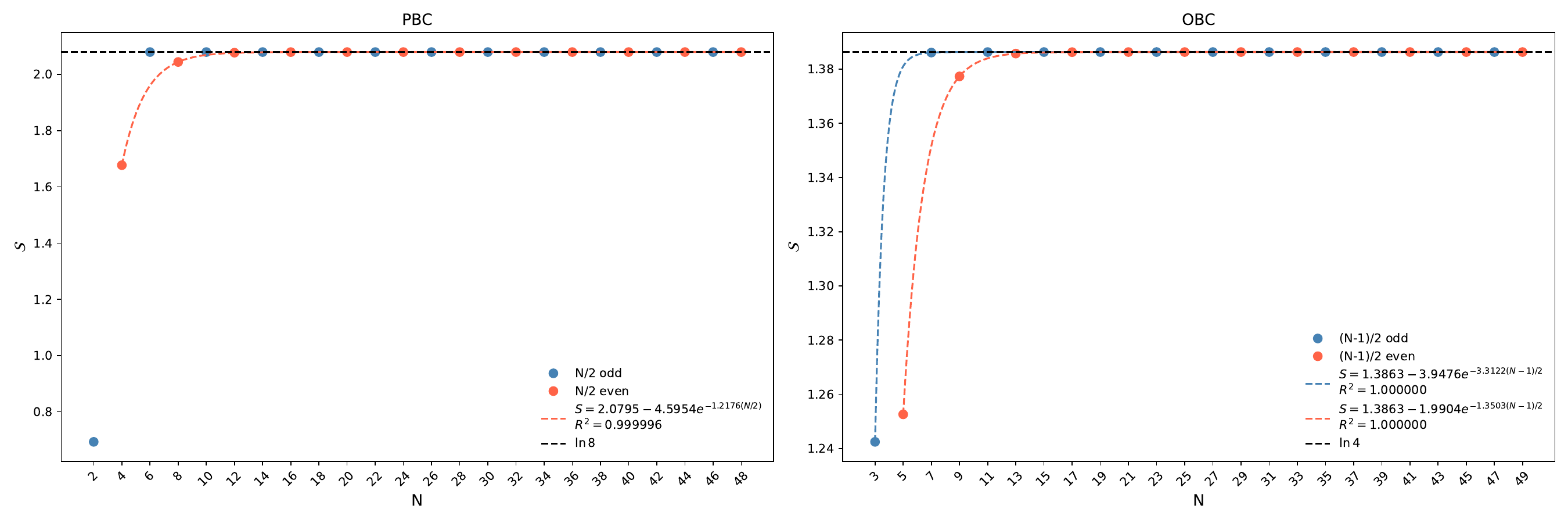}
\caption{Half-chain entanglement entropy of the FD phase as a function of system size $N$. Left: even $N$ 
under PBC, with the $N/2$-odd sub-series (blue) sitting exactly on $\ln 8 = 3\ln 2$ and the $N/2$-even 
sub-series (orange) fitted to Eq.~\eqref{eq:fit_form}. Right: odd $N$ under OBC, with both sub-series fitted 
to Eq.~\eqref{eq:fit_form} and converging to $\ln 4 = 2\ln 2$ (dashed line).}
\label{fig8}
\end{figure}

Figure~\ref{fig8} shows the half-chain entropy of the FD phase under both boundary conditions. 
Under PBC the $N/2$-odd sub-series attains $S_\infty = \ln 8 = 3\ln 2$ exactly from small $N$ with no 
finite-size correction, while the $N/2$-even sub-series rises steeply from below and converges rapidly to the 
same limit. The saturation $\ln 8$ arises from three virtual spin-$\tfrac{1}{2}$ singlets crossing the cut, 
each contributing $\ln 2$. Under OBC both sub-series converge to $\ln 4 = 2\ln 2$, with the odd sub-series 
converging noticeably faster (larger $C$, Table~\ref{tab:fd_halfchain_fit}), reflecting the reduced 
contribution of the free edge spin-$\tfrac{3}{2}$ at the boundary.

\begin{table}[h]
\centering
\caption{Fit parameters of Eq.~\eqref{eq:fit_form} for the FD phase.}
\label{tab:fd_halfchain_fit}
\begin{tabular}{llcccc}
\hline\hline
BC & Sub-series & $S_\infty$ & $B$ & $C$ & $R^2$ \\
\hline
PBC & $N/2$ even & $\ln 8 \approx 2.0795$ & $4.5954$ & $1.2176$ & $0.9999$ \\
PBC & $N/2$ odd & $\ln 8 \approx 2.0794$ & \multicolumn{3}{c}{exact} \\
\hline
OBC & $(N-1)/2$ odd  & $\ln 4 \approx 1.3863$ & $3.9476$ & $3.3122$ & $\approx 1.0000$ \\
OBC & $(N-1)/2$ even & $\ln 4 \approx 1.3863$ & $1.9904$ & $1.3503$ & $\approx 1.0000$ \\
\hline\hline
\end{tabular}
\end{table}

\paragraph{Partially dimerized phase.}
 
\begin{figure}[h]
\centering
\includegraphics[width=\textwidth]{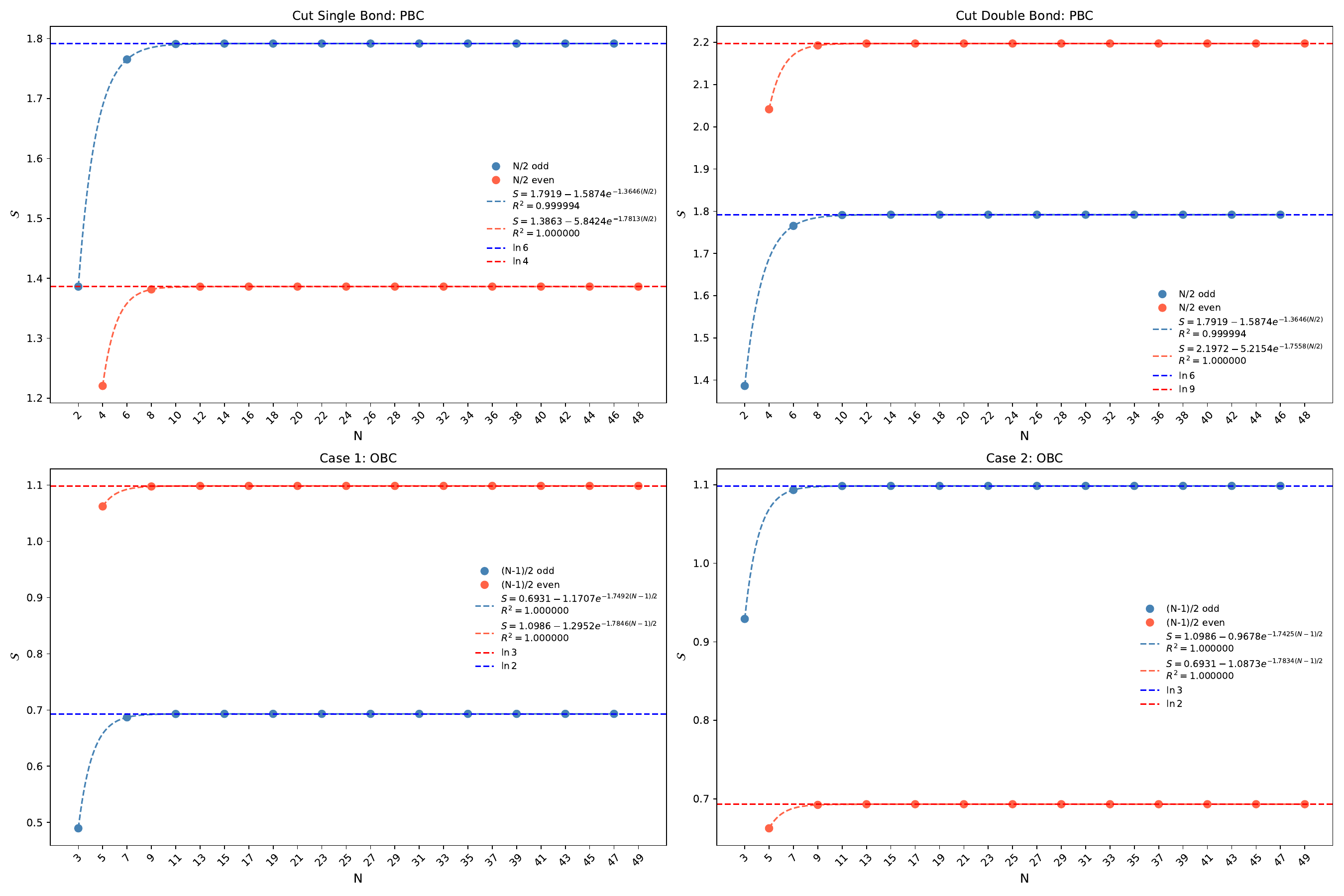}
\caption{Half-chain entanglement entropy of the PD phase as a function of system size $N$. Top row: even $N$ 
under PBC for a single-bond cut (left) and a double-bond cut (right), with the $N/2$-odd sub-series (blue) 
converging to $\ln 6$ and the $N/2$-even sub-series (orange) converging to $\ln 4$ or $\ln 9$ respectively. 
Bottom row: odd $N$ under OBC for Case~1 left=single (left) and Case~2 left=double (right), with each sub 
series converging to $\ln 2$ or $\ln 3$ depending on the subsystem size parity. All curves are fitted to 
Eq.~\eqref{eq:fit_form} with parameters listed in
Tables~\ref{tab:pd_pbc_fit} and~\ref{tab:pd_obc_fit}.}
\label{fig9}
\end{figure}

Figure~\ref{fig9} shows the half-chain entropy of the PD phase across all four boundary 
configurations. The PD phase exhibits multiple distinct saturation values depending on both the boundary 
condition and the bond type at the cut. Under PBC (top row), both cut types share the same $N/2$-odd 
saturation $S_\infty = \ln 6 = \ln 2 + \ln 3$, arising from a mixed Schmidt structure where one single and 
one double virtual leg cross the cut simultaneously. The $N/2$-even sub-series, however, splits according to 
bond type: the single-bond cut saturates to $\ln 4 = 2\ln 2$, while the double-bond cut saturates to 
$\ln 9 = 2\ln 3$, with the even sub-series rising steeply from below in both cases. Under OBC (bottom row), 
the saturation is governed by the edge configuration rather than the cut type: Case~1 (left=single) has the 
odd sub-series converging to $\ln 2$ and the even sub-series to $\ln 3$, while Case~2 (left=double) shows the 
reversed pattern, with both cases converging rapidly within $N \approx 10$. All OBC sub-series share 
$C \approx 1.75$--$1.78$ (Table~\ref{tab:pd_obc_fit}), indicating a uniform correlation length regardless of 
the boundary configuration. The coexistence of $\ln 2$, $\ln 3$, $\ln 4$, $\ln 6$, and $\ln 9$ as saturation 
values within the same phase is the definitive fingerprint of partial dimerization, absent in both the MG 
model and the FD phase where all cuts are equivalent.

\begin{table}[h]
\centering
\caption{Fit parameters of Eq.~\eqref{eq:fit_form} for the PD phase under PBC.}
\label{tab:pd_pbc_fit}
\begin{tabular}{llcccc}
\hline\hline
Cut type & Sub-series & $S_\infty$ & $B$ & $C$ & $R^2$ \\
\hline
Single bond & $N/2$ odd  & $\ln 6 \approx 1.7918$ & $1.5874$ & $1.3646$ & $0.9999$ \\
Single bond & $N/2$ even & $\ln 4 \approx 1.3863$ & $5.8424$ & $1.7813$ & $\approx 1.0000$ \\
\hline
Double bond & $N/2$ odd  & $\ln 6 \approx 1.7918$ & $1.5874$ & $1.3646$ & $0.9999$ \\
Double bond & $N/2$ even & $\ln 9 \approx 2.1972$ & $5.2154$ & $1.7558$ & $\approx 1.0000$ \\
\hline\hline
\end{tabular}
\end{table}

\begin{table}[h]
\centering
\caption{Fit parameters of Eq.~\eqref{eq:fit_form} for the PD phase under OBC.}
\label{tab:pd_obc_fit}
\begin{tabular}{llcccc}
\hline\hline
Case & Sub-series & $S_\infty$ & $B$ & $C$ & $R^2$ \\
\hline
Case 1 (left=single) & $(N-1)/2$ odd  & $\ln 2 \approx 0.6931$ & $1.1707$ & $1.7492$ & $\approx 1.0000$ \\
Case 1 (left=single) & $(N-1)/2$ even & $\ln 3 \approx 1.0986$ & $1.2952$ & $1.7846$ & $\approx 1.0000$ \\
\hline
Case 2 (left=double) & $(N-1)/2$ odd  & $\ln 3 \approx 1.0986$ & $0.9678$ & $1.7425$ & $\approx 1.0000$ \\
Case 2 (left=double) & $(N-1)/2$ even & $\ln 2 \approx 0.6931$ & $1.0873$ & $1.7834$ & $\approx 1.0000$ \\
\hline\hline
\end{tabular}
\end{table}

The MG model (spin-$\tfrac{1}{2}$) and the FD phase (spin-$\tfrac{3}{2}$) share the same fully dimerized 
ground-state structure, differing only in spin magnitude and the number of singlets per bond. This is directly 
reflected in the half-chain entropy: under PBC the saturation follows $S_\infty = n\ln 2$, where $n$ is the 
number of virtual spin-$\tfrac{1}{2}$ singlets crossing the cut --- one singlet per bond for the MG model 
gives $S_\infty = 2\ln 2$, while three singlets per bond for the FD phase give $S_\infty = \ln 8 = 3\ln 2$. 
Under OBC the free edge spin reduces $S_\infty$ by one $\ln 2$ contribution: the MG edge spin-$\tfrac{1}{2}$ 
loses one virtual leg giving $S_\infty = \tfrac{3}{2}\ln 2$, while the FD edge spin-$\tfrac{3}{2}$ likewise 
loses one virtual leg out of three giving $S_\infty = \ln 4 = 2\ln 2$. Both models also exhibit the same 
qualitative even--odd oscillation pattern and exponential finite-size convergence, with the FD phase 
converging faster (larger $C$) owing to the stronger singlet bonds in the spin-$\tfrac{3}{2}$ chain. 
Crucially, the finite saturation of $S_\infty$ to a constant independent of system size $N$ in both models 
confirms compliance with the area law for gapped one-dimensional systems~\cite{Hastings2007,Eisert2010}: the 
entanglement entropy of a subsystem is bounded by a constant determined solely by the number of singlets 
crossing the boundary, not by the subsystem volume. The same area-law saturation is observed in the PD phase, 
where $S_\infty$ likewise converges to a finite constant for every boundary configuration, confirming that 
partial dimerization does not alter the area-law nature of the entanglement despite the coexistence of 
inequivalent bond types.

\subsection{Single-Site Entanglement Entropy}

\subsubsection{Majumdar--Ghosh Model}
 
\begin{figure}[h]
\centering
\includegraphics[width=\textwidth]{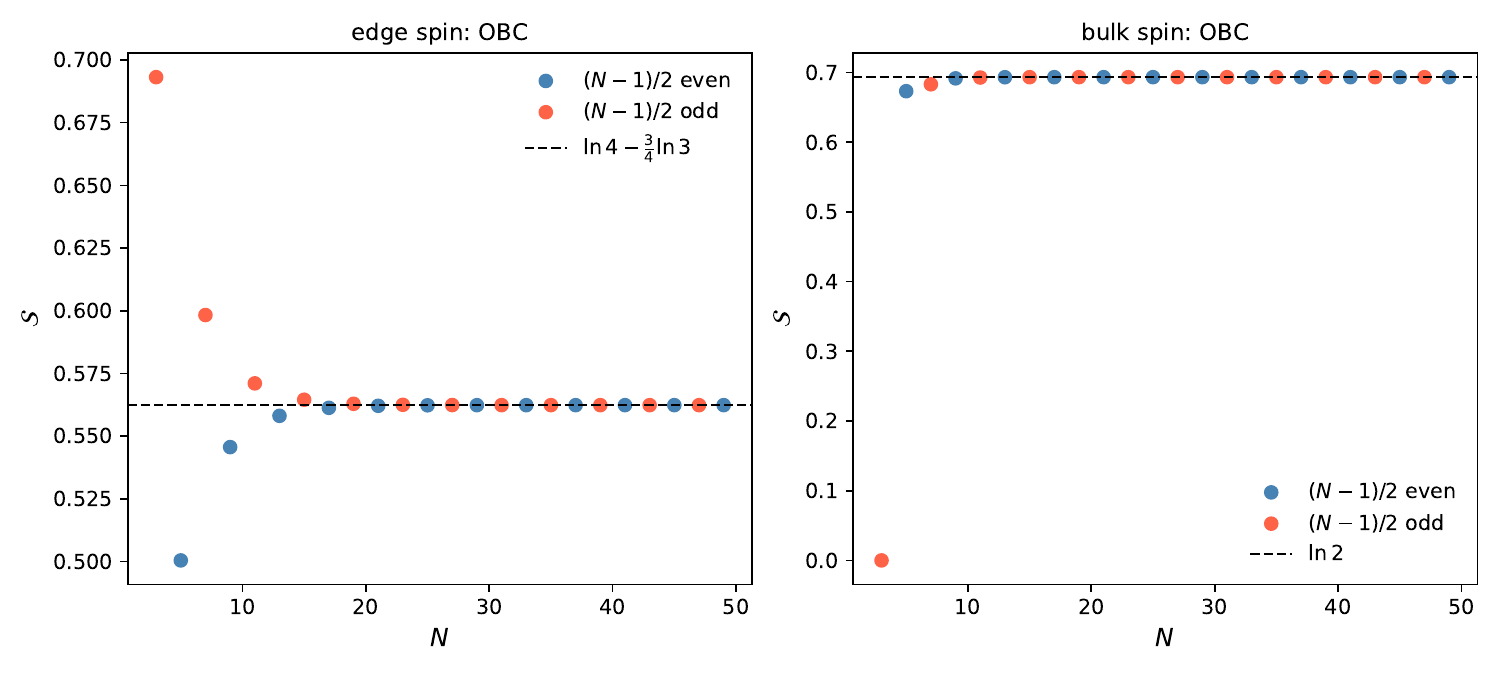}
\caption{Single-site entanglement entropy of the MG model for odd $N$ under OBC as a function of system size 
$N$. Left: edge site $s=0$, with both sub-series converging from opposite sides to 
$\ln 4 - \tfrac{3}{4}\ln 3$ (dashed line). Right: bulk site $s=1$, with both sub-series converging
rapidly to $\ln 2$ (dashed line), except at $N=3$ where $S(s=1)=0$ exactly.}
\label{fig10}
\end{figure}

Figure~\ref{fig10} shows the single-site entropy at the edge site $s=0$ and the bulk site $s=1$ as 
functions of $N$ for odd chains under OBC. For the bulk site (right panel), both sub series converge rapidly to

\begin{equation}
    S_{\text{bulk}}^{\text{MG}} = \ln 2 \approx 0.6931,
    \label{eq:mg_bulk_single}
\end{equation}
with the exception of $N=3$, where $S(s=1) = 0$ exactly because the middle site forms no singlet bond and its 
reduced state fully factorizes from the rest of the chain. The bulk value $\ln 2$ reflects that every interior 
spin-$\tfrac{1}{2}$ participates in exactly one nearest-neighbor singlet, leaving its reduced density matrix 
as an equal mixture of the two spin states. Under PBC with even $N$, all sites attain $S(s) = \ln 2$ exactly 
and uniformly for all accessible $N$, with no finite-size correction.
 
For the edge site (left panel), the two sub-series converge from opposite sides to the exact analytical value

\begin{equation}
    S_{\text{edge}}^{\text{MG}}
    = \ln 4 - \tfrac{3}{4}\ln 3 \approx 0.5623,
    \label{eq:mg_edge_single}
\end{equation}
with the odd sub-series approaching from above and the even sub-series from below, both converging by 
$N \approx 20$. This value is suppressed relative to $\ln 2$ due to the partially free boundary 
spin-$\tfrac{1}{2}$ that is only weakly entangled through the surrounding dimer background.

\subsubsection{Spin-$\tfrac{3}{2}$ $J_1$--$J_2$--$J_3$ Chain}

\paragraph{Fully dimerized phase.}
 
\begin{figure}[h]
\centering
\includegraphics[width=\textwidth]{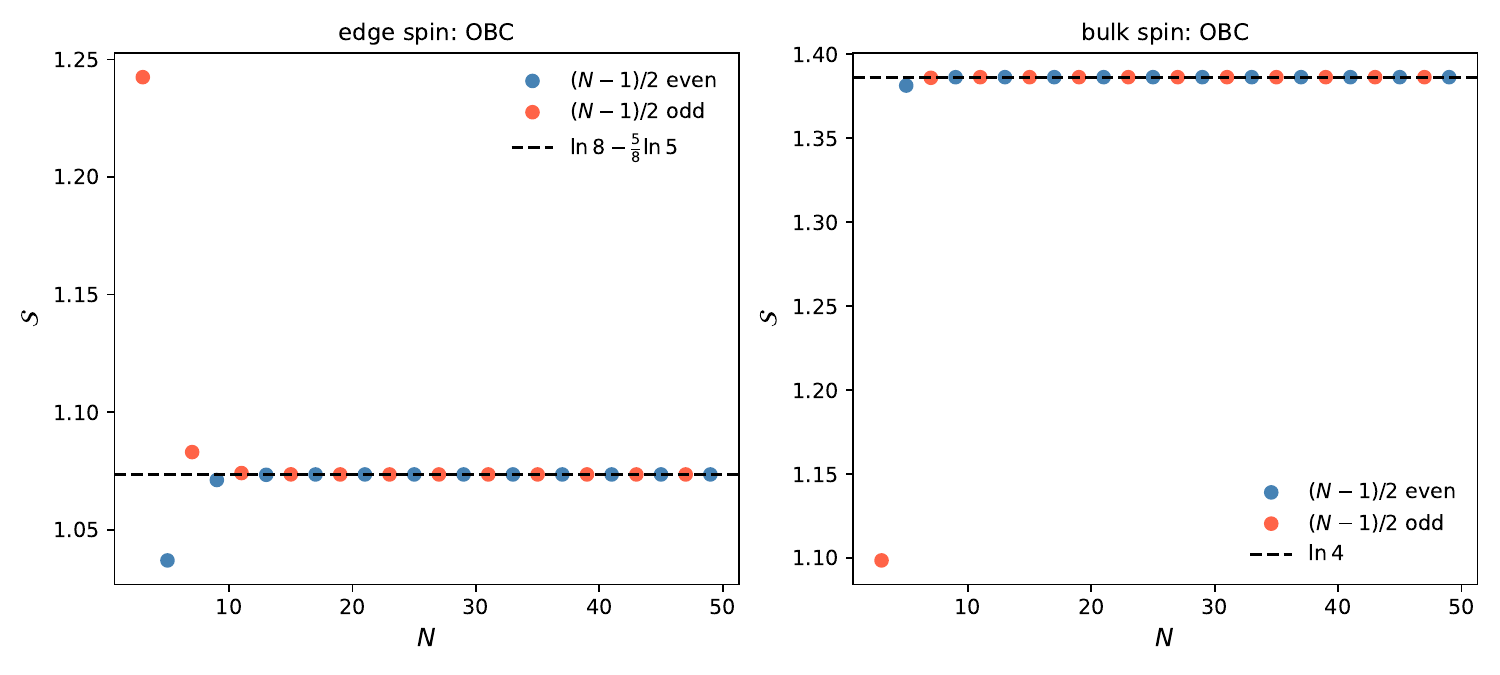}
\caption{Single-site entanglement entropy of the FD phase for odd $N$ under OBC as a function of system size 
$N$. Left: edge spin, with the odd sub-series (orange) approaching from above and the even sub-series (blue) 
from below, both converging to $\ln 8 - \tfrac{5}{8}\ln 5$ (dashed line). Right: bulk spin, with both 
sub-series converging rapidly to $\ln 4$ (dashed line) from below, except at $N=3$ where the odd sub-series 
starts at $\approx 1.099$.}
\label{fig11}
\end{figure}

Figure~\ref{fig11} shows the edge and bulk single-site entropy for the FD phase under OBC with odd 
$N$. For the bulk spin (right panel), both sub-series converge from below to

\begin{equation}
    S_{\text{bulk}}^{\text{FD}} = \ln 4 = 2\ln 2 \approx 1.3863,
    \label{eq:fd_bulk_single}
\end{equation}
the maximum possible value for a spin-$\tfrac{3}{2}$ ($d=4$) site, indicating that every bulk spin is 
maximally entangled with its environment. Both sub-series are nearly indistinguishable and converge rapidly by 
$N \approx 10$. Under PBC with even $N$, all sites attain $S(s) = \ln 4$ exactly and uniformly for all 
accessible $N$, with no finite-size correction.
 
For the edge spin (left panel), the odd sub-series starts high at $N=3$ ($\approx 1.242$) and converges from 
above, while the even sub-series starts low at $N=5$ ($\approx 1.037$) and converges from below, both reaching 
the exact analytical value

\begin{equation}
    S_{\text{edge}}^{\text{FD}}
    = \ln 8 - \tfrac{5}{8}\ln 5 \approx 1.0735,
    \label{eq:fd_edge_single}
\end{equation}
by $N \approx 15$. This value is suppressed relative to $\ln 4$ due to the free spin-$\tfrac{3}{2}$ degree of 
freedom localized at the boundary under OBC with odd $N$. Comparing with the MG model, both edge values follow 
the same pattern of suppression below the bulk maximum, with the FD edge value $\ln 8 - \tfrac{5}{8}\ln 5$ 
playing the same role as $\ln 4 - \tfrac{3}{4}\ln 3$ in the MG case.

\paragraph{Partially dimerized phase.}
 
The single-site entropy of the PD phase is richer than the FD phase due to the inequivalent boundary 
configurations. Under PBC with even $N$, all sites attain $S(s) = \ln 4$ exactly and uniformly for all 
accessible $N$, identical to the FD phase, with translational symmetry masking the alternating bond structure 
entirely.
 
\begin{figure}[h]
\centering
\includegraphics[width=\textwidth]{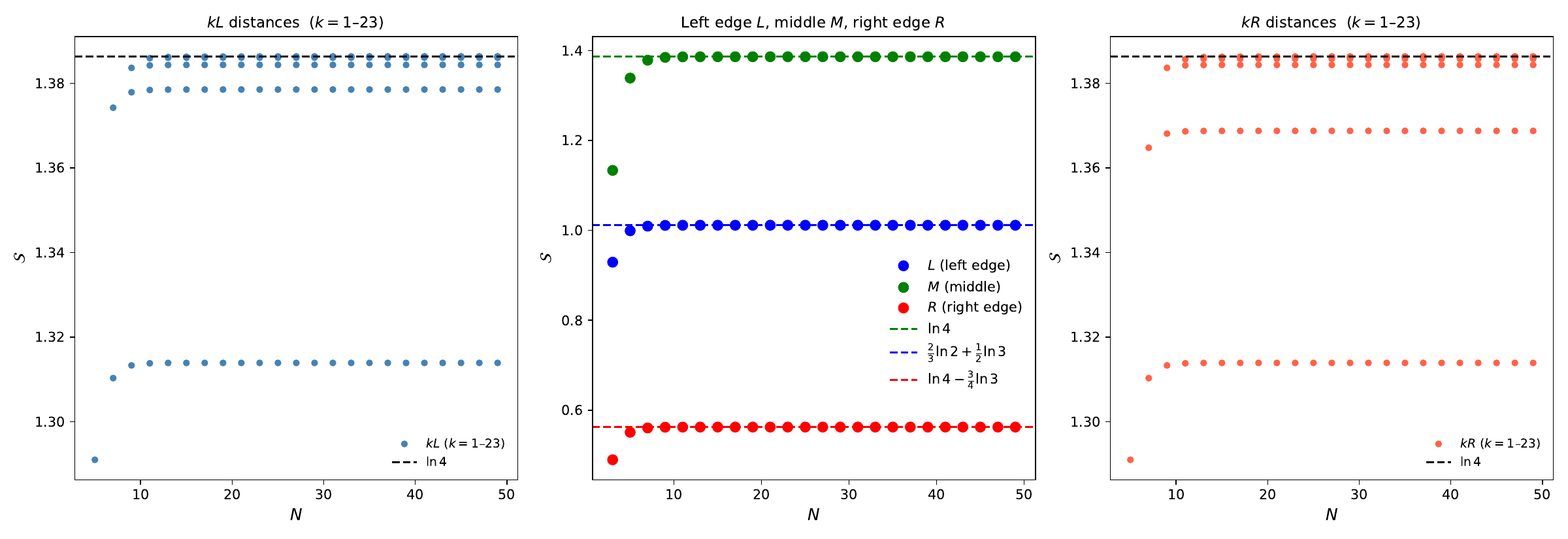}
\caption{Single-site entanglement entropy of the PD phase for odd $N$ under OBC (left=single configuration) as 
a function of system size $N$. Left: interior sites $kL$ ($k=1$--$23$), all converging to $\ln 4$ (dashed) 
via a staggered two-band structure. Middle: left edge $L$ (blue), middle $M$ (green), and right edge $R$ 
(red), converging to $\tfrac{2}{3}\ln 2 + \tfrac{1}{2}\ln 3$, $\ln 4$, and $2\ln 2 - \tfrac{3}{4}\ln 3$ 
respectively (dashed lines). Right: interior sites $kR$ ($k=1$--$23$), converging to $\ln 4$ (dashed) via the 
same two-band structure as $kL$.}
\label{fig12}
\end{figure}

Under OBC with odd $N$, we label the left edge spin $L$, the right edge spin $R$, the middle spin $M$, and 
denote the spin at distance $k$ from the left (right) edge as $kL$ ($kR$). Figure~\ref{fig12} 
shows the convergence of $S(L)$, $S(M)$, $S(R)$, and the $kL$/$kR$ bands as functions of $N$ for the 
left=single configuration. The middle spin (green) saturates to the maximum value

\begin{equation}
    S_M = \ln 4 \approx 1.3863,
    \label{eq:pd_bulk_single}
\end{equation}
identical to the FD bulk, already from small $N$. The two edge spins converge to distinct exact values 
reflecting the asymmetry of the boundary:

\begin{align}
    S_L &= \tfrac{2}{3}\ln 2 + \tfrac{1}{2}\ln 3 \approx 1.008,
    \label{eq:pd_SL} \\
    S_R &= 2\ln 2 - \tfrac{3}{4}\ln 3 \approx 0.563,
    \label{eq:pd_SR}
\end{align}
where $S_L > S_R$ because the left edge carries a single virtual spin-$\tfrac{1}{2}$ while the right edge 
carries a different virtual configuration with weaker entanglement. The interior sites $kL$ and $kR$ (left and 
right panels) each display a staggered two-band structure in which odd-$k$ and even-$k$ sites approach $\ln 4$ 
along two distinct trajectories, converging by $N \approx 30$. Beyond this boundary layer all interior sites 
are indistinguishable from the FD bulk value $\ln 4$, confirming that single-site entropy cannot distinguish 
the FD and PD phases in the bulk.

\subsection{Pairwise Entanglement Entropy}
 
\subsubsection{Majumdar--Ghosh Model}

\begin{figure}[h]
\centering
\includegraphics[width=0.7\textwidth]{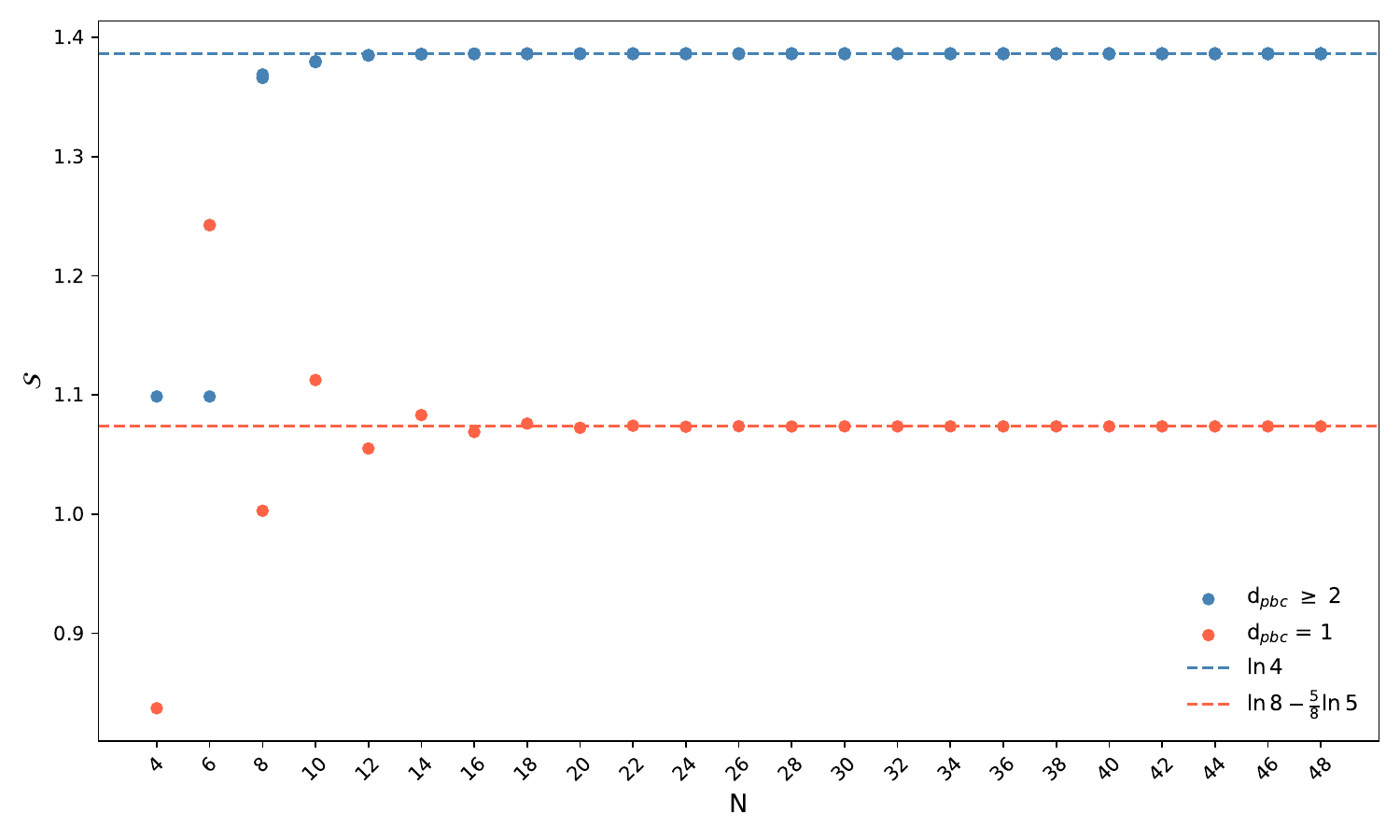}
\caption{Pairwise entanglement entropy of the MG model for even $N$ under PBC. Nearest-neighbor pairs 
($d_{pbc}=1$, orange) converge to $3\ln 2 - \tfrac{5}{8}\ln 5$, while all longer-distance pairs 
($d_{pbc}\geq 2$, blue) saturate rapidly to $2\ln 2$.}
\label{fig13}
\end{figure}

\paragraph{Periodic boundary conditions (even $N$)}
Figure~\ref{fig13} shows the pairwise entropy for even $N$ under PBC as a function of $N$. The 
nearest neighbor pairs ($d_{pbc}=1$, orange) converge from below to

\begin{equation}
    S(d_{pbc}=1) = 3\ln 2 - \tfrac{5}{8}\ln 5 \approx 1.0735,
    \label{eq:mg_pbc_nn}
\end{equation}
while all longer-distance pairs ($d_{pbc} \geq 2$, blue) saturate rapidly to

\begin{equation}
    S(d_{pbc}\geq 2) = 2\ln 2 \approx 1.3863,
    \label{eq:mg_pbc_bulk}
\end{equation}
independent of distance, already for $N \geq 6$. The reduced nearest-neighbor value arises because the two 
adjacent sites share a singlet bond, consuming part of their entanglement with the environment. The identical 
saturation of $S(d_{pbc}\geq 2)$ to $2\ln 2$ under both PBC and OBC (bulk--bulk) confirms that the bulk 
pairwise entanglement structure is boundary-independent in the thermodynamic limit.

\begin{figure}[h]
\centering
\includegraphics[width=0.7\textwidth]{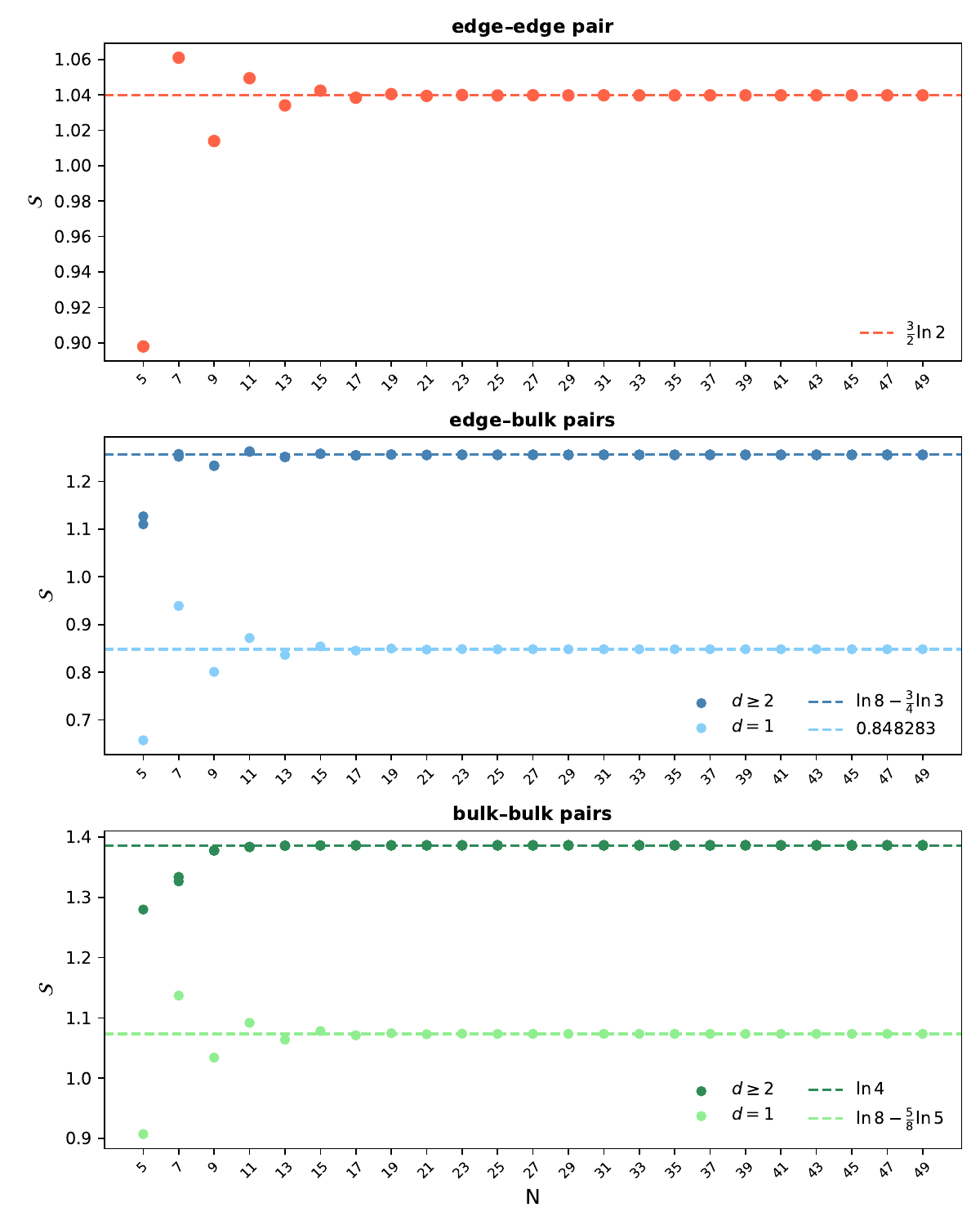}
\caption{Pairwise entanglement entropy of the MG model for odd $N$ under OBC. Top: edge--edge pair converging 
to $\tfrac{3}{2}\ln 2$. Middle: edge--bulk pairs with $d=1$ (light blue) converging to $\approx 0.848283$ and 
$d\geq 2$ (dark blue) converging to $3\ln 2 - \tfrac{3}{4}\ln 3$. Bottom: bulk--bulk pairs with $d=1$ (light 
green) converging to $\ln 8 - \tfrac{5}{8}\ln 5$ and $d\geq 2$ (dark green) converging to $2\ln 2$.}
\label{fig14}
\end{figure}

\paragraph{Open boundary conditions (odd $N$)}
Figure~\ref{fig14} shows the pairwise entropy for odd $N$ under OBC, classified into three 
groups. For the \textit{edge--edge pair} (top panel), the single data series converges to

\begin{equation}
    S_{ee} = \tfrac{3}{2}\ln 2 \approx 1.0397,
    \label{eq:mg_ee}
\end{equation}
matching the half-chain OBC saturation value, with small oscillations before settling by $N \approx 15$. For 
the \textit{edge--bulk pairs} (middle panel), the nearest-neighbor pair ($d=1$, left edge paired with first 
bulk spin) saturates to $\approx 0.848283$, while all longer-distance pairs ($d \geq 2$) saturate to

\begin{equation}
    S_{eb}(d\geq 2) = 3\ln 2 - \tfrac{3}{4}\ln 3 \approx 1.2408.
    \label{eq:mg_eb}
\end{equation}
For the \textit{bulk--bulk pairs} (bottom panel), the nearest-neighbor pair ($d=1$) saturates to the exact 
value

\begin{equation}
    S_{bb}(d=1) = \ln 8 - \tfrac{5}{8}\ln 5 \approx 1.0735,
    \label{eq:mg_bb_nn}
\end{equation}
while all longer-distance bulk--bulk pairs ($d \geq 2$) saturate to

\begin{equation}
    S_{bb}(d\geq 2) = 2\ln 2 \approx 1.3863,
    \label{eq:mg_bb}
\end{equation}
independent of distance. The saturation of $S_{bb}(d\geq 2)$ to a distance-independent constant confirms that 
two bulk spins not sharing a singlet bond are entangled with the rest of the chain in an identical fashion 
regardless of their separation, a direct signature of the short-range product-of-singlets structure.

\subsubsection{Spin-$\tfrac{3}{2}$ $J_1$--$J_2$--$J_3$ Chain}
 
\paragraph{Fully dimerized phase}
\begin{figure}[h]
\centering
\includegraphics[width=0.7\textwidth]{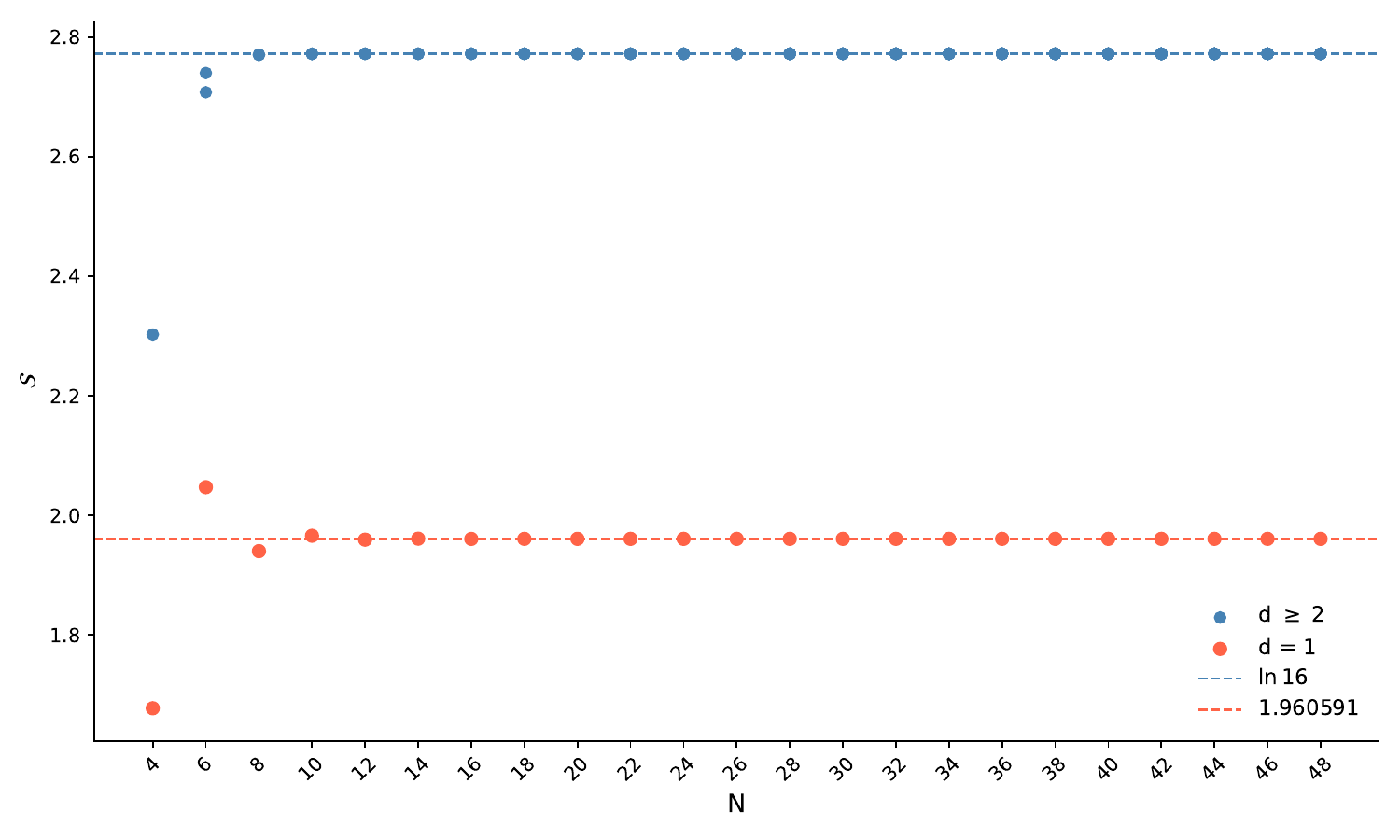}
\caption{Pairwise entanglement entropy of the FD phase for even $N$ under PBC. Nearest-neighbor pairs ($d=1$, 
orange) converge to $1.960591$, while all longer-distance pairs ($d\geq 2$, blue) saturate rapidly to 
$\ln 16 = 4\ln 2$.}
\label{fig15}
\end{figure}
 
\begin{figure}[h]
\centering
\includegraphics[width=0.7\textwidth]{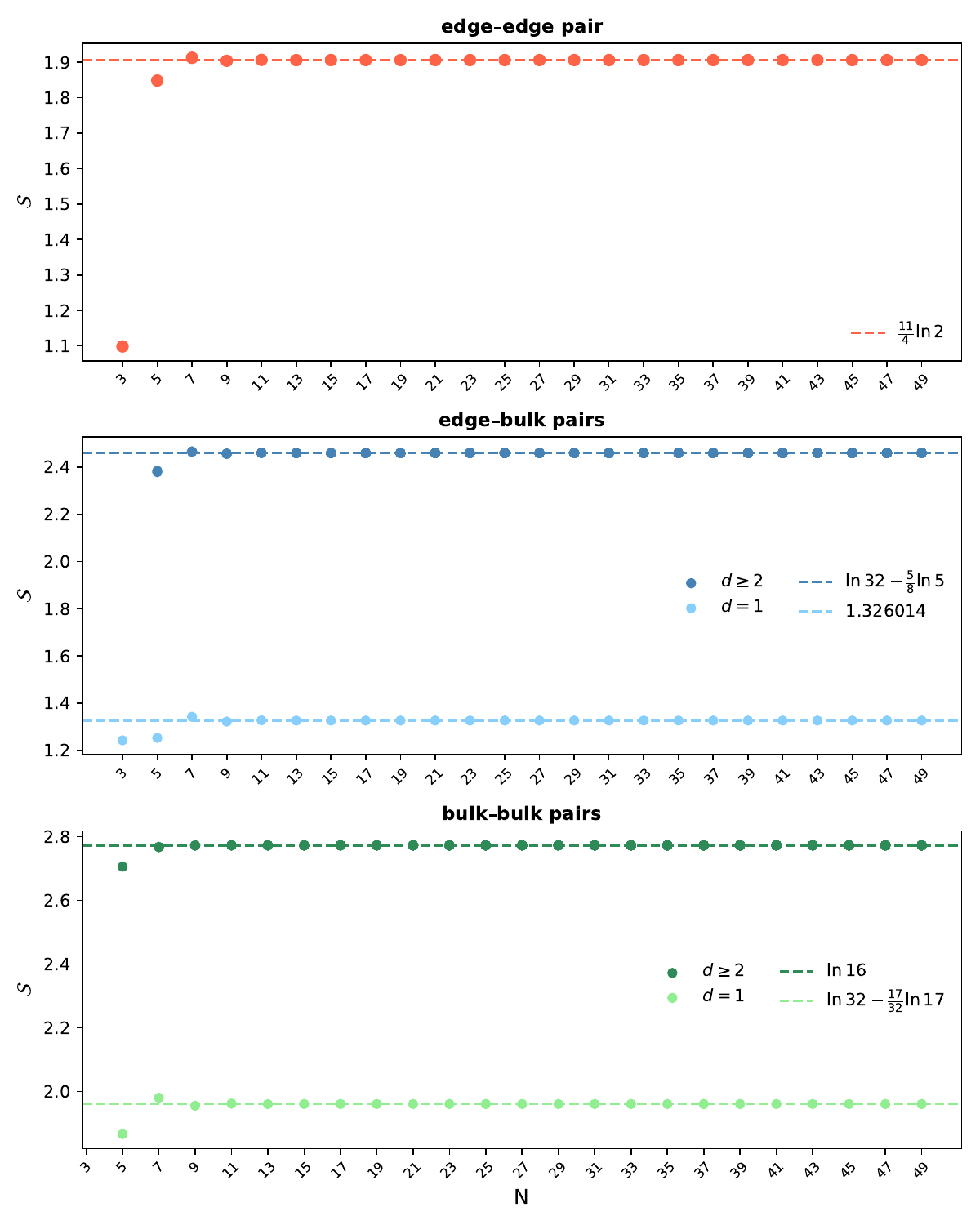}
\caption{Pairwise entanglement entropy of the FD phase for odd $N$ under OBC. Top: edge--edge pair converging 
to $\tfrac{11}{4}\ln 2$. Middle: edge--bulk pairs with $d=1$ (light blue) converging to $1.326014$ and 
$d\geq 2$ (dark blue) converging to $\ln 32 - \tfrac{5}{8}\ln 5$. Bottom: bulk--bulk pairs with $d=1$ 
(light green) converging to $\ln 32 - \tfrac{17}{32}\ln 17$ and $d\geq 2$ (dark green) converging to 
$\ln 16 = 4\ln 2$.}
\label{fig16}
\end{figure}
 
Figure~\ref{fig15} and~\ref{fig16} show the pairwise entropy of the FD phase under PBC and OBC respectively. 
Under PBC (Fig.~\ref{fig15}), the nearest neighbor pairs ($d=1$) converge to $1.960591$, while all 
longer-distance pairs ($d \geq 2$) saturate rapidly to

\begin{equation}
    S_{bb}^{\text{FD}}(d\geq 2) = \ln 16 = 4\ln 2 \approx 2.7726,
    \label{eq:fd_bb}
\end{equation}
independent of distance, already for small $N$.
 
Under OBC (Fig.~\ref{fig16}), the three pair groups yield distinct saturation values. 
The \textit{edge--edge pair} converges rapidly to

\begin{equation}
    S_{ee}^{\text{FD}} = \tfrac{11}{4}\ln 2 \approx 1.9054,
    \label{eq:fd_ee}
\end{equation}
already by $N \approx 7$. For \textit{edge--bulk pairs}, the nearest-neighbor pair ($d=1$) saturates 
to $1.326014$, while all longer-distance pairs ($d \geq 2$) converge to

\begin{equation}
    S_{eb}^{\text{FD}}(d\geq 2) = \ln 32 - \tfrac{5}{8}\ln 5 \approx 2.4327.
    \label{eq:fd_eb}
\end{equation}
For \textit{bulk--bulk pairs}, the nearest-neighbor pair ($d=1$) saturates to

\begin{equation}
    S_{bb}^{\text{FD}}(d=1) = \ln 32 - \tfrac{17}{32}\ln 17 \approx 1.9606,
    \label{eq:fd_bb_nn}
\end{equation}
while all longer-distance bulk--bulk pairs ($d \geq 2$) saturate to $\ln 16$, identical to the PBC value, 
confirming boundary-independence of the bulk pairwise entropy in the thermodynamic limit.

\paragraph{Partially dimerized phase}
\begin{figure}[h]
\centering
\includegraphics[width=0.6\textwidth]{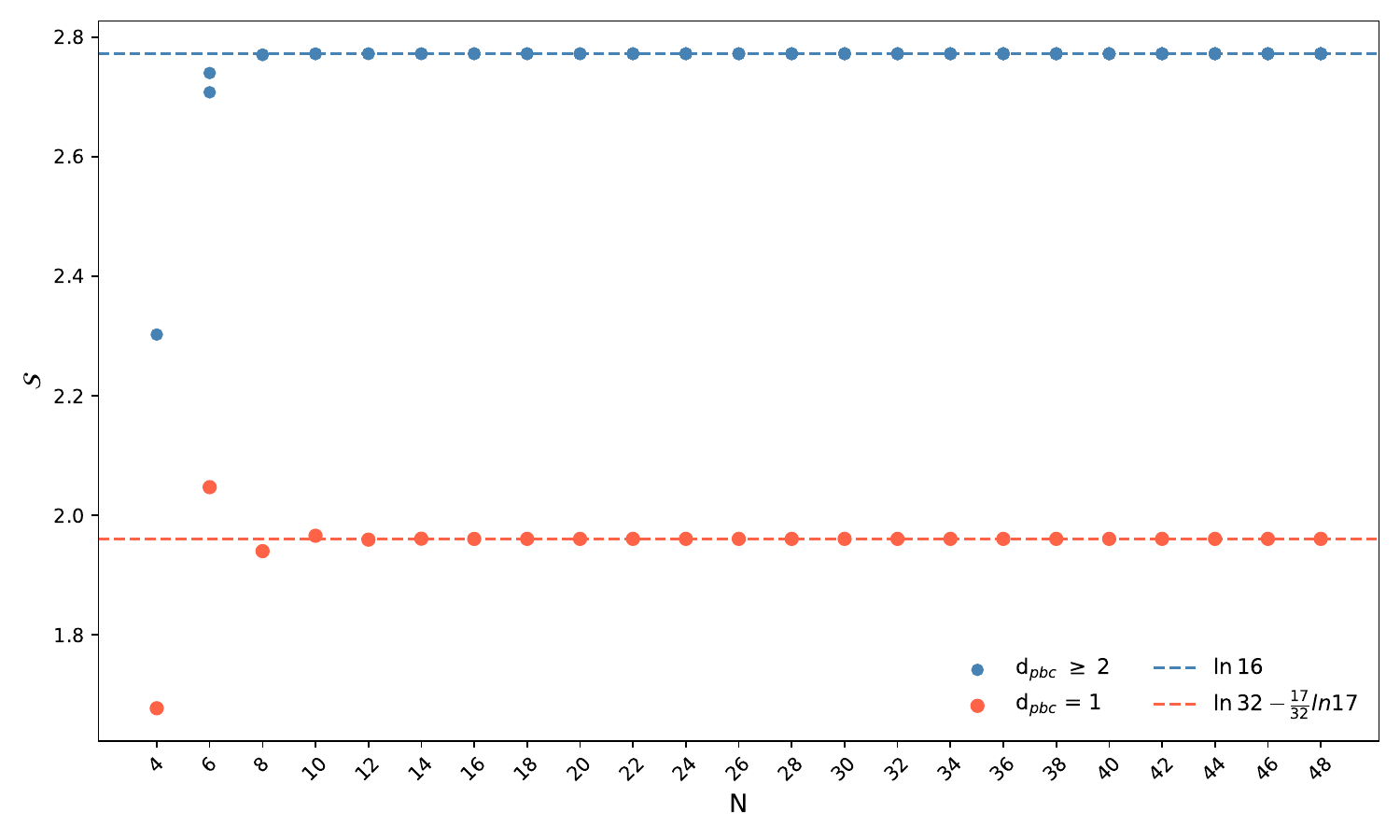}
\caption{Pairwise entanglement entropy of the PD phase for even $N$ under PBC. Nearest-neighbor pairs 
($d_{pbc}=1$, orange) converge to $\ln 32 - \tfrac{17}{32}\ln 17$, while all longer-distance 
pairs ($d_{pbc}\geq 2$, blue) saturate to $\ln 16 = 4\ln 2$.}
\label{fig17}
\end{figure}
 
\begin{figure}[h]
\centering
\includegraphics[width=\textwidth]{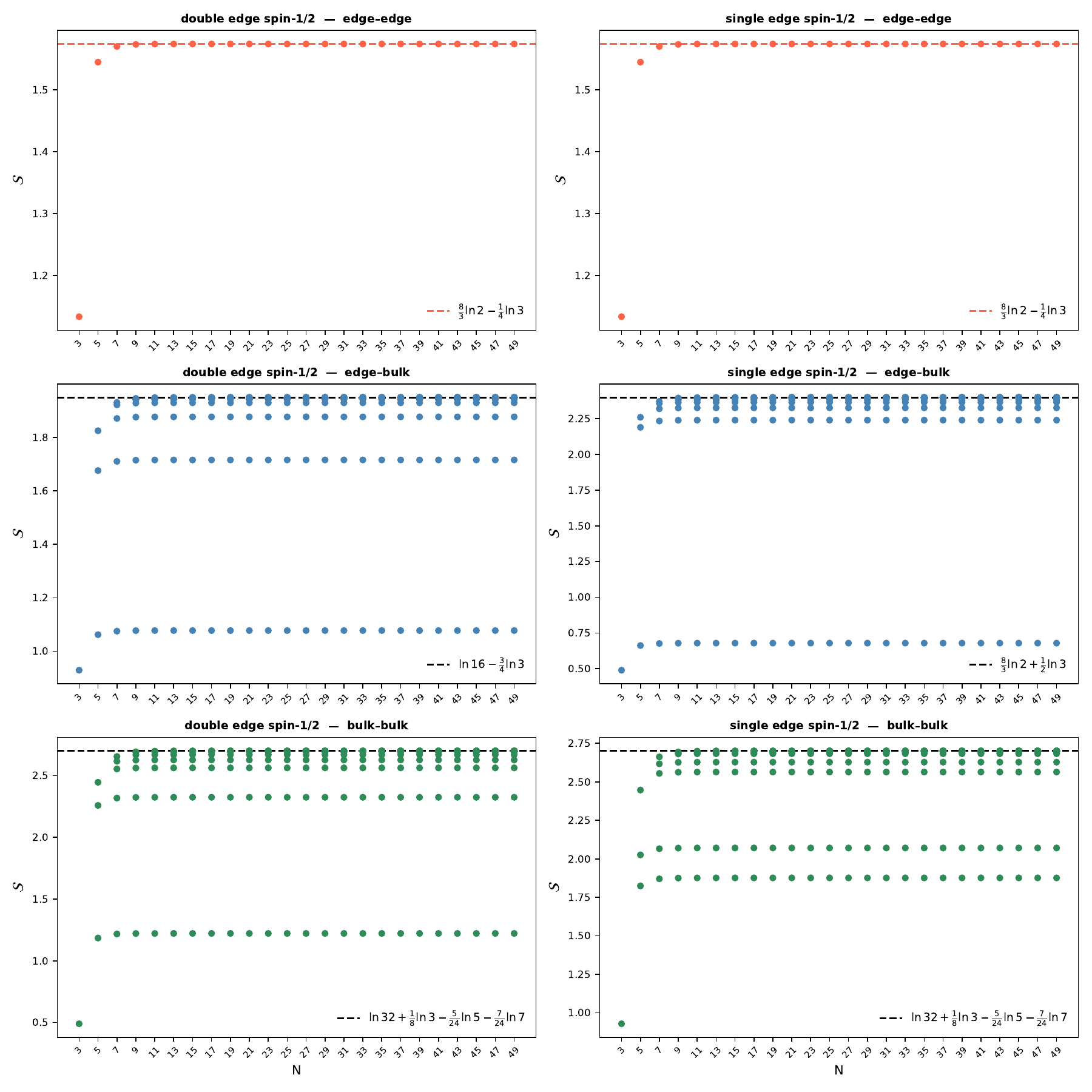}
\caption{Pairwise entanglement entropy of the PD phase for odd $N$ under OBC, for double edge 
spin-$\tfrac{1}{2}$ (left column) and single edge spin-$\tfrac{1}{2}$ (right column). In each column, rows 
show edge--edge (top), edge--bulk (middle), and bulk--bulk (bottom) pairs. Within each panel the bands are 
ordered from $d=1$ (bottom) to larger $d$ (above), converging to their respective dashed-line limits.}
\label{fig18}
\end{figure}
 
Figure~\ref{fig17} and~\ref{fig18} show the pairwise entropy of the PD phase. Under PBC (Fig.~\ref{fig17}), 
the nearest-neighbor pairs ($d_{pbc}=1$) converge to

\begin{equation}
    S^{\text{PD}}(d_{pbc}=1) = \ln 32 - \tfrac{17}{32}\ln 17 \approx 1.9606,
    \label{eq:pd_pbc_nn}
\end{equation}
while all longer-distance pairs ($d_{pbc}\geq 2$) saturate to $\ln 16 = 4\ln 2 \approx 2.7726$, identical to 
the FD phase under PBC.
 
Under OBC (Fig.~\ref{fig18}), the results depend on the boundary configuration but the three pair groups share 
the same structure in both cases. For both double and single edge spin-$\tfrac{1}{2}$ configurations, the 
\textit{edge-edge pair} converges to the same value

\begin{equation}
    S_{ee}^{\text{PD}} = \tfrac{8}{3}\ln 2 - \tfrac{1}{4}\ln 3,
    \label{eq:pd_ee}
\end{equation}
reflecting the fact that the edge--edge entropy depends only on the two boundary spins regardless of which 
configuration is chosen. For \textit{edge--bulk pairs}, the two configurations differ: the double edge gives a 
top-band limit of $\ln 16 - \tfrac{3}{4}\ln 3$, while the single edge gives $\tfrac{8}{3}\ln 2 + 
\tfrac{1}{2}\ln 3$, with the bands ordered from $d=1$ at the bottom rising to the respective limits. 
For \textit{bulk--bulk pairs}, both configurations share the same top-band saturation

\begin{equation}
    S_{bb}^{\text{PD}}(d\geq 2) = \ln 32 + \tfrac{1}{8}\ln 3
    - \tfrac{5}{24}\ln 5 - \tfrac{7}{24}\ln 7,
    \label{eq:pd_bb}
\end{equation}
with lower bands descending from $d=1$ at the bottom. The multiple distinct bands in the edge-bulk and 
bulk-bulk panels of the PD phase, absent in the FD phase, are the most direct local signature of the 
alternating bond structure: each distance $d$ probes a different combination of single- and double-singlet 
bonds along the chain, producing a rich multi-band pairwise entropy spectrum unique to the PD phase.

\section{Conclusion}
\label{sec:conclusion}

We determine the von Neumann entanglement entropy of exact valence-bond ground states in the spin-$\tfrac{1}{2}$ Majumdar--Ghosh 
model and the spin-$\tfrac{3}{2}$ $J_1$--$J_2$--$J_3$ chain, covering fully and partially dimerized phases across half-chain, 
single-site, and pairwise bipartitions under both open and periodic boundary conditions. In all cases, the entropy saturates to a 
finite constant with increasing system size $N$, establishing area-law behavior~\cite{Hastings2007,Eisert2010} and confirming 
entanglement entropy as a controlled probe of frustrated dimerized spin systems.

The MG model and the fully dimerized phase exhibit the same entanglement structure: even--odd oscillations in half-chain scaling, 
exponential approach to saturation, edge suppression under open boundaries, and uniform bulk single-site entropy. Their 
differences are purely quantitative, set by spin magnitude and bond multiplicity. By contrast, the partially dimerized phase 
displays distinct and robust signatures: multiple saturation values in the half-chain entropy determined by the bond configuration 
at the cut, asymmetric edge single-site entropies, and a pronounced multi-band structure in the pairwise entropy, where each 
separation resolves a specific combination of single and double singlet bonds. This multi-band structure is absent in fully 
dimerized phases and provides a definitive entanglement fingerprint of partial dimerization.

These results establish entanglement entropy computed from exact ground states as a quantitative and discriminating diagnostic of 
frustrated dimerized phases, directly encoding their bond architecture beyond the reach of conventional order parameters or 
phase-boundary analyses.

\begin{acknowledgments}
This research was supported by the Research Grant, Faculty of Science, Prince of Songkla University (contract no. SCITUG65001).
\end{acknowledgments}

\bibliography{references}

\appendix

\section{Matrix representation}
\label{appdxA}

We represent each physical spin-$\frac{3}{2}$ as a fully symmetric combination of three virtual spin-$\frac{1}{2}$ degrees of freedom. 
For a given physical projection $m \in \{-\frac{3}{2}, -\frac{1}{2}, \frac{1}{2}, \frac{3}{2}\}$, we derived
\begin{equation}
\begin{split}
T^{[m]}&=\sum_{m=i+j+k}C^{ijk}|s^i_1\rangle \otimes |s^j_2\rangle \otimes |s^k_3\rangle\\
       &=\sum_{m=i+j+k} C^{ijk}|m\rangle,
\end{split}
\end{equation}
where \( |s\rangle\in \left\{ |\uparrow\rangle=\begin{bmatrix}
1 \\ 0
\end{bmatrix}, |\downarrow\rangle=\begin{bmatrix}
0 \\ 1
\end{bmatrix} \right\} \) , $i,j,k = \pm\frac{1}{2}$, and $C^{ijk}$ are Clebsch-Gordan coefficients for the symmetric spin-$\frac{3}{2}$ sector. In the chosen basis, the spin-$\frac{3}{2}$ vectors read
\[
\begin{aligned}
T^{\left[\frac{3}{2}\right]} &=
\begin{bmatrix}
\left|\frac{3}{2}\right\rangle \\ 0 \\ 0 \\ 0 \\ 0 \\ 0 \\ 0 \\ 0
\end{bmatrix},\quad
T^{\left[\frac{1}{2}\right]} &=
\frac{1}{\sqrt{3}}
\begin{bmatrix}
0 \\ \left|\frac{1}{2}\right\rangle \\ \left|\frac{1}{2}\right\rangle \\ 0 \\ \left|\frac{1}{2}\right\rangle \\ 0 \\ 0 \\ 0
\end{bmatrix},\quad
T^{\left[-\frac{1}{2}\right]} &=
\frac{1}{\sqrt{3}}
\begin{bmatrix}
0 \\ 0 \\ 0 \\ \left|-\frac{1}{2}\right\rangle \\ 0 \\ \left|-\frac{1}{2}\right\rangle \\ \left|-\frac{1}{2}\right\rangle \\ 0
\end{bmatrix},\quad
T^{\left[-\frac{3}{2}\right]} &=
\begin{bmatrix}
0 \\ 0 \\ 0 \\ 0 \\ 0 \\ 0 \\ 0 \\ \left|-\frac{3}{2}\right\rangle
\end{bmatrix}.
\end{aligned}
\]
To form singlet bonds on each virtual leg, we apply the singlet matrix $\Phi_{ss}$ to every spin-$\frac{1}{2}$ vector. 

\begin{equation}
\Phi_{ss} = \frac{1}{\sqrt{2}}\begin{bmatrix}
0 & 1 \\
-1 & 0 \\
\end{bmatrix}.
\end{equation}
This defines modified spin-$\frac{3}{2}$ vectors $\tilde{T}^{[m]}$

\begin{equation}
\begin{split}
\tilde{T}^{[m]}&=\sum_{m=i+j+k}C^{ijk} \Phi_{ss}|s^i_1 \rangle \otimes \Phi_{ss}|s^j_2\rangle \otimes \Phi_{ss}|s^k_3\rangle \\
       &=\sum_{m=i+j+k} \tilde{C}^{ijk}|m\rangle,
\end{split}
\end{equation}
where $\tilde{C}^{ijk}$ the renormalized coefficients. In the same basis,
\[
\begin{aligned}
\tilde{T}^{\left[\frac{3}{2}\right]} &=
-\frac{1}{2\sqrt{2}}
\begin{bmatrix}
0 \\ 0 \\ 0 \\ 0 \\ 0 \\ 0 \\ 0 \\ \left|\tfrac{3}{2}\right\rangle
\end{bmatrix},\quad
\tilde{T}^{\left[\frac{1}{2}\right]} &=
\frac{1}{2\sqrt{6}}
\begin{bmatrix}
0 \\ 0 \\ 0 \\ \left|\tfrac{1}{2}\right\rangle \\ 0 \\ \left|\tfrac{1}{2}\right\rangle \\ \left|\tfrac{1}{2}\right\rangle \\ 0
\end{bmatrix},\quad
\tilde{T}^{\left[-\frac{1}{2}\right]} &=
-\frac{1}{2\sqrt{6}}
\begin{bmatrix}
0 \\ \left|-\tfrac{1}{2}\right\rangle \\ \left|-\tfrac{1}{2}\right\rangle \\ 0 \\ \left|-\tfrac{1}{2}\right\rangle \\ 0 \\ 0 \\ 0
\end{bmatrix},\quad
\tilde{T}^{\left[-\frac{3}{2}\right]} &=
\frac{1}{2\sqrt{2}}
\begin{bmatrix}
\left|-\tfrac{3}{2}\right\rangle \\ 0 \\ 0 \\ 0 \\ 0 \\ 0 \\ 0 \\ 0
\end{bmatrix}.
\end{aligned}
\]

\section{Rationale for the exponential fit}
\label{appdxB}

In one-dimensional spin systems, two functional forms are commonly considered for the finite-size scaling of the half-chain 
entanglement entropy. The first is the exponential form
\begin{equation}
    S(\ell) = A - B\,e^{-C\ell},
    \label{eq:fit_exp}
\end{equation}
which describes a gapped system obeying the area law~\cite{Hastings2007,Eisert2010}, where the entropy saturates to a finite 
constant $A = S_\infty$ and the finite-size corrections decay exponentially at a rate $C = 1/\xi$ governed by the bulk 
correlation length $\xi$. The second is the logarithmic form
\begin{equation}
    S(\ell) = \frac{c}{6}\ln\ell + \mathrm{const},
    \label{eq:fit_log}
\end{equation}
which describes a critical gapless system governed by conformal field theory, where $c$ is the central charge~\cite{Calabrese2004,
Vidal2003}. Choosing between these two forms is not merely a numerical fitting decision — it carries direct physical meaning.
Exponential saturation identifies the ground state as gapped and non-critical, while logarithmic growth signals criticality.
 
\begin{figure}[h]
    \centering
    \includegraphics[width=\textwidth]{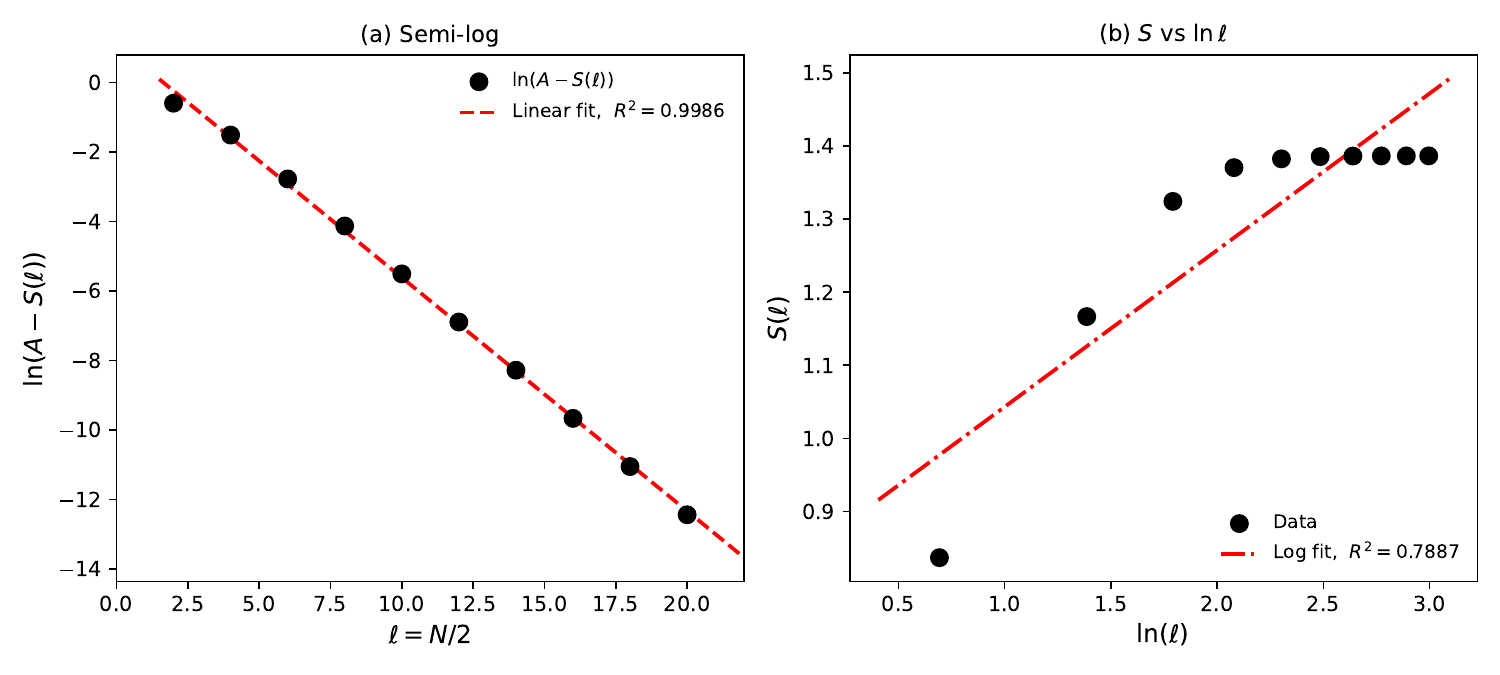}
    \caption{Diagnostic test of the exponential fitting form using the $N/2$-even sub-series of the half-chain entanglement entropy of 
    the Majumdar--Ghosh model under PBC. (a) Semi-logarithmic plot of $\ln(A - S(\ell))$ versus $\ell = N/2$, confirming exponential 
    convergence ($R^2 = 0.9986$). (b) Plot of $S(\ell)$ versus $\ln\ell$, showing that logarithmic scaling is ruled out 
    ($R^2 = 0.7887$).}
    \label{fig:fit_example}
\end{figure}

To determine which form applies to the systems studied in this work, the half-chain entanglement entropy of the $N/2$-even 
sub-series of the Majumdar--Ghosh model under periodic boundary conditions is chosen as a representative test case. This example 
captures the typical scaling behavior observed across the datasets considered here. Both fitting forms are applied to this data, 
and the corresponding diagnostic plots are presented in Fig.~\ref{fig:fit_example}.

In panel~(a), $\ln(A - S(\ell))$ is plotted as a function of $\ell = N/2$ on a semi-logarithmic scale. For exponential 
convergence, this quantity is expected to exhibit linear behavior. As seen in the figure, the data points closely follow a straight 
line, with a fit quality of $R^2 = 0.9986$, providing strong evidence that the entanglement entropy converges exponentially to 
its saturation value. In contrast, panel~(b) shows $S(\ell)$ plotted against $\ln \ell$ to test for logarithmic scaling, which would 
appear as a linear relation in this representation. The data, however, display clear deviations from linearity, and the corresponding 
fit yields a significantly lower coefficient of determination, $R^2 = 0.7887$. This comparison indicates that logarithmic scaling 
does not adequately describe the data.
 
The comparison between the two panels establishes that the half-chain entropy of the Majumdar--Ghosh model under periodic 
boundary conditions is well described by an exponential form rather than a logarithmic one. This behavior is consistent with the 
fully dimerized, gapped nature of the Majumdar--Ghosh ground state, which is characterized by a finite correlation length set by 
the singlet bond structure. The absence of logarithmic scaling further indicates that the system does not exhibit conformal field 
theory behavior associated with critical spin chains. Motivated by this result, the exponential form in Eq.~\eqref{eq:fit_exp} 
is adopted as the fitting function for all half-chain entropy sub-series throughout this work.

\end{document}